\documentclass[a4paper,aps,pre,showpacs,superscriptaddress,groupedaddress,twocolumn]{revtex4}

\usepackage{amsmath}
\usepackage{amssymb}
\usepackage{graphicx}
\usepackage{xcolor}
\usepackage{natbib}
\usepackage{siunitx}
\usepackage{appendix}

\usepackage{ulem}
\newcommand{\beq}{\begin{equation}}
\newcommand{\eeq}{\end{equation}}

\newcommand{\upd}{\mathrm{d}}	

\newcommand{\gbar}{\bar{\gamma}}  

\newcommand{\FPressure}{F_P}
\newcommand{\FTension}{F_T}

\newcommand{\BVPAngle}{\varphi}
\newcommand{\h}{\bar{h}}
\newcommand{\Dh}{\Delta h}
\newcommand{\ho}{h_0}
\newcommand{\R}{R}
\newcommand{\asp}{\alpha}
\newcommand{\Vtot}{V_{\text{tot}}}
\newcommand{\Ftot}{F_{\text{tot}}}
\newcommand{\Smax}{\varsigma}
\newcommand{\smax}{s_{\text{max}}}
\newcommand{\Fflat}{F_{\text{flat}}}
\newcommand{\Reql}{a_0}
\newcommand{\Ca}{\mbox{Ca}}
\newcommand{\Fmax}{F_{\text{max}}}
\newcommand{\V}{V_1}
\newcommand{\F}{F_1}
\newcommand{\REQL}{\hat{a}_0}
\newcommand{\Rpert}{A}
\newcommand{\Ppert}{P}
\newcommand{\hdimless}{\hat{h}}
\newcommand{\rdimless}{\hat{r}}
\newcommand{\Vdimless}{\hat{V}}
\newcommand{\pdimless}{\hat{p}}
\newcommand{\xdimless}{\hat{x}}

\newcommand{\BVPr}{\rho}
\newcommand{\BVPz}{\zeta}
\newcommand{\BVPR}{\varrho}
\newcommand{\BVPZ}{\xi}
\newcommand{\nbridge}{\mathcal{N}}
\newcommand{\Fbridge}{\mathcal{F}}

\begin{document}

\title{Liquid bridge splitting enhances normal capillary adhesion and resistance to shear on rough surfaces}

\author{Matthew D. Butler} 
\affiliation{Mathematical Institute, University of Oxford, Woodstock Rd, Oxford, OX2 6GG, United Kingdom}
\affiliation{School of Mathematics, University of Birmingham, Edgbaston, B15 2TT, United Kingdom}
\author{Dominic Vella}
\affiliation{Mathematical Institute, University of Oxford, Woodstock Rd, Oxford, OX2 6GG, United Kingdom}


\date{\today}

\begin{abstract}
The effect of `bridge splitting' is considered in the case of capillary adhesion: for a fixed total volume of liquid, does having more capillary bridges increase the total adhesion force?
Previous studies have shown that the capillary-induced adhesion force between two planar surfaces is only substantially enhanced by bridge splitting in specific circumstances. Here this previous result is reconsidered, and it is shown that bridge splitting may significantly increase the adhesion forces when one of the surfaces is rough. The resistance to shear is also examined, and it is shown that bridge splitting on a rough surface can lead to a steady capillary-induced shear force that scales linearly with translation velocity, even in the absence of contact-line pinning.
\end{abstract}


\maketitle

\section{Introduction}


The ability of some animals to climb vertical surfaces and walk upside-down has been an area of scientific interest for hundreds of years \cite{Hooke1665,Power1664,Autumn2002,Hanna1991,Ishii1987}. One particularly diverse group of climbing animals are insects. Across a large range of body sizes and masses \cite{Labonte2015}, insects are able to reliably stick to and walk on a variety of surfaces \cite{England2016} even under large externally applied loads. It is believed that insect adhesion is aided by the effect of the surface tension of an oily secretion beneath their feet \cite{Dirks2011review} --- insects adhere, at least in part, due to capillary effects.

Capillary adhesion is also prevalent in many other commonplace and technological settings: in everyday life the condensation on the outside of a glass of a cold drink may cause a coaster to remain attached to the bottom of the glass as it is lifted, while the grains in a sandcastle are best cohered at an optimal volume fraction of liquid \cite{Halsey1998,Pakpour2012} (with similar considerations being important in understanding the cohesion of soil \cite{Haines1925,Fisher1926,Middleton1994}). In technological applications, the presence of capillary bridges can be both problematic, such as in the stiction of magnetic storage discs \cite{Mate1992,Gao1995} and also during the production of micro-electro-mechanical systems (MEMS) \cite{Tanaka1993,Mastrangelo1993}, but can also be beneficial, such as helping to form nano-scale structured assemblies during the drying of pre-wetted nanotube forests \cite{Chakrapani2004,Pokroy2009,deVolder2013}.

The wide range of scenarios involving liquids bridging two solid surfaces has motivated many previous modelling studies  of such systems, beginning with the first studies of capillarity by \citet{Young1805} and Laplace \cite{Laplace1806}. The shape of a liquid bridge has been studied for many different combinations of solid surface geometry, allowing the adhesive force to be calculated: in particular, solutions exist for the shape of a bridge between two rigid plates \cite{Carter1988}, two spheres \cite{Melrose1966,Willett2000} and a sphere-plate combination \cite{Orr1975}. Moreover, studies of stability and rupture explain when these bridges may break \cite{Lowry1995}. 

The details of capillary bridge shape are quite involved for general bridge sizes. However, when the bridge is relatively wide compared to its height (i.e.~when the surfaces are very close to one another), the problem simplifies considerably because the meniscus shape is approximately a circular arc \cite{Fisher1926}. The pressure within the bridge is then inversely proportional to the separation between the surfaces at the contact line; for liquids that wet the surfaces sufficiently, the pressure within the bridge is a suction and provides the dominant adhesion force. This simplification allows for approximate analytical calculations to be performed for a variety of geometries \cite{Butt2009,Cai2008}, and shows that the capillary adhesion force between smooth planar surfaces is proportional to the total volume of liquid in the bridge \cite{Meurisse2006} (as we shall see in \S\ref{sec:SimpleAdhesion}). However, many observational studies show that some insects (including flies and beetles) have hairy footpads, which give many points of adhesive contact \cite{Gorb1998,Eisner2000}, whilst others (including some species of ants and stick insects) have an adhesive consisting of a water-in-oil emulsion: many small water droplets are dispersed in an oil \cite{Federle2002,Dirks2010}. 

The ubiquity of some form of  `bridge splitting' in natural occurrences of capillary adhesion suggests that it must  be somehow beneficial \cite{Dirks2011review}. An obvious possibility is that such an emulsion somehow enhances capillary adhesion, but this appears to be at odds with the conclusion from the simple models that the capillary adhesion force is simply proportional to liquid volume; in this case adhesion would be unchanged by splitting a single bridge into a large number of smaller bridges with the same total volume. 
However, the simplified solutions with adhesion force proportional to volume do not quite give the whole story --- the details of the bridge shape can make a significant difference in the adhesion force, particularly when liquid bridges have a width comparable to their height. 
A more detailed study of the effect of dividing a fixed volume of liquid into many bridges between flat plates was presented by \citet{DeSouza2008split}.  They found that, with a fixed liquid volume, it is indeed possible to increase the adhesion force by having many bridges, but that this enhancement is modest (giving less than a two-fold increase in adhesion, as we shall quantify later) when the surface is well-wetted by the liquid (as is believed to be the case in many physiologically relevant cases). If the bridge splitting observed in insect adhesion does not lead to a sizeable increase in the adhesion force on smooth surfaces then the question becomes: what is its purpose?

One key piece of physics that was omitted in the splitting study of \citet{DeSouza2008split} is surface roughness, which is known to play an important role in adhesion. For `dry' adhesives on rough surfaces, there can be a trade-off between adhesive attraction due to, for example, van der Waals forces, and repulsion from deformation of the adhering solid: if the surface is sufficiently rough, or the adhering solid sufficiently stiff, then the surfaces may not be able to conform to one another, resulting in only partial adhesion \cite{Zilberman2002,Persson2003}.  Perhaps to combat this, many climbing organs have a large number of small contacts \cite{Federle2006}. For dry adhesion, the theory of \citet{Johnson1971} (often called JKR adhesion) suggests that there is a finite pull-off force that scales linearly with with the radius of spherical contacts. When dividing a single large pad into many smaller pads, we may then expect an increase in the total adhesion force for a similar contact area \cite{Arzt2003}, though this may be limited by other physical constraints \cite{Spolenak2005}. Having many smaller contacts may be additionally beneficial by allowing the adhesive organ to conform more closely to a rough substrate \cite{Beutel2001}. 

In this article we address the question of whether bridge splitting is similarly beneficial for capillary adhesion to a rough surface. One important feature of capillary adhesion that is qualitatively different to the dry case is that neither the number nor the position of capillary bridges are necessarily fixed. Indeed, when surfaces are in close contact in a humid environment, condensation can form new capillary bridges between them, and it has been shown that the roughness of these surfaces can play a significant role in the resulting adhesion \cite{Wang2009,DelRio2007,Persson2008,Rabinovich2002}. 

\begin{figure}[tbp]
	\centering
	\includegraphics[width=\linewidth]{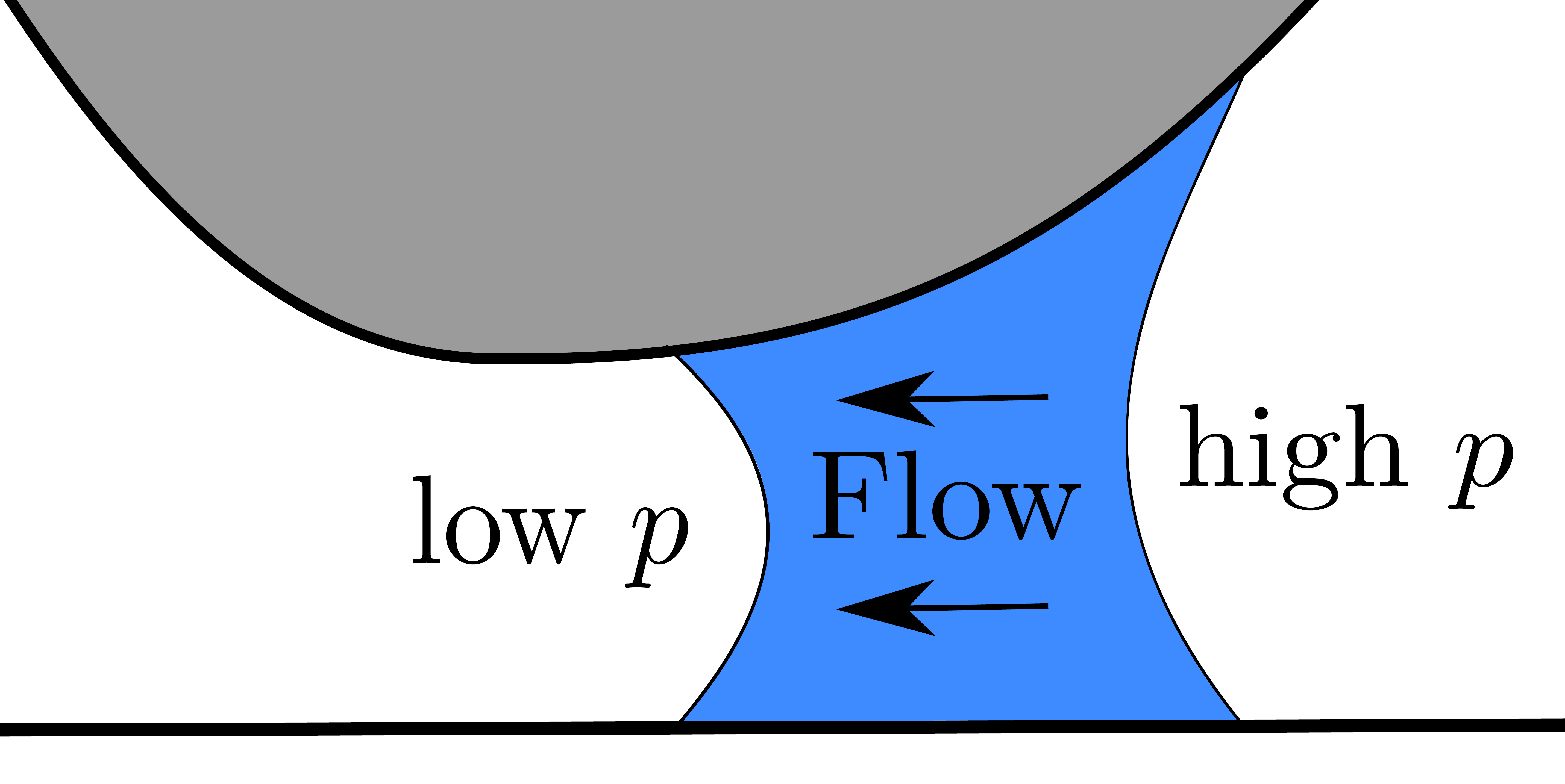}
	\caption{Wetting liquid bridges migrate towards local minima in gap separation. At the narrower end of the bridge, its surface is more curved, resulting in a capillary pressure gradient inside the liquid that induces an internal flow towards the narrow end.}
	\label{fig:DropMigration}
\end{figure}

\begin{figure*}[tbp]
	\begin{center}
		\includegraphics[width=\linewidth]{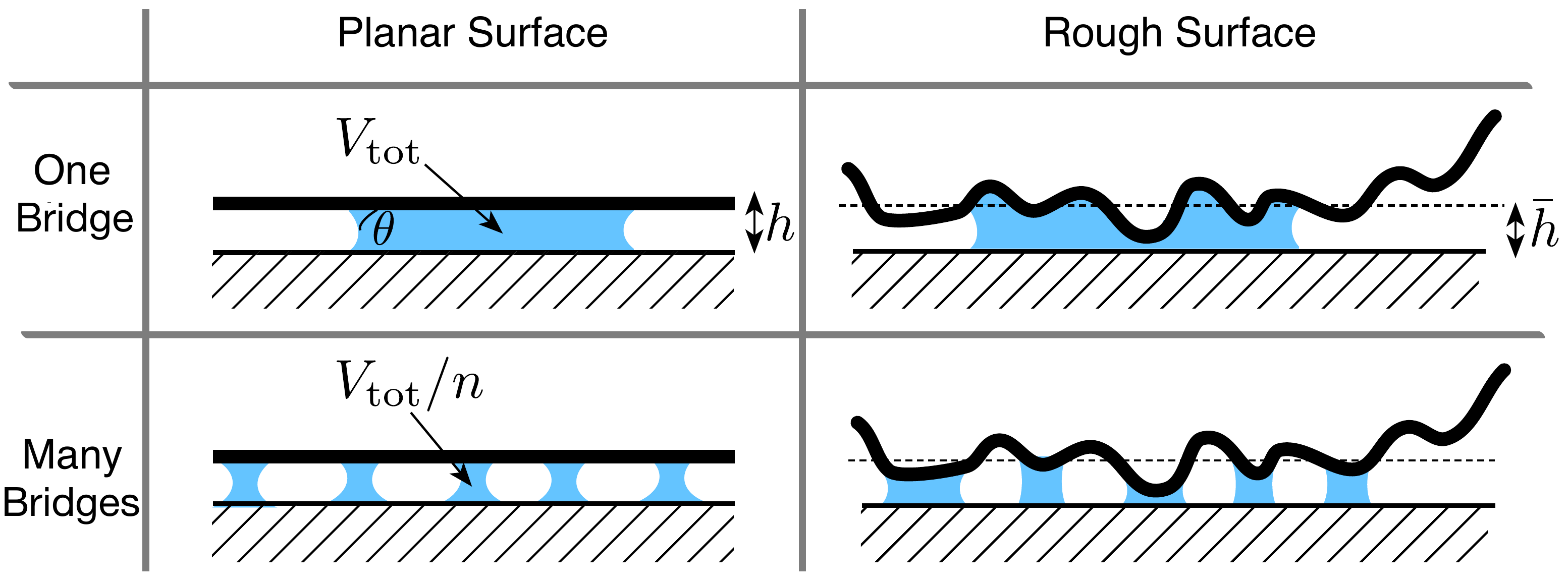} 
		\caption[Bridge splitting on planar and rough surfaces]{Capillary adhesion between planar (left) and rough (right) surfaces, as mediated by a single bridge (top) and many, smaller, bridges (bottom). In each case, the total volume of liquid is $\Vtot$, which may be enclosed within a single bridge, or within many. The contact angle of the liquid is shown as the angle $\theta$ and the separation of the surfaces is $h$ in the smooth case, while for rough surfaces an equivalent parameter is the mean separation, $\bar{h}$.} 
	\label{fig:schematic} 
	\end{center}
\end{figure*}

The possibility of capillary condensation is well-studied, but is unlikely to have a direct impact in cases where controlled adhesion is required, such as the adhesion of insects, since adhesion must occur regardless of external humidity. However, the significance of capillary bridge mobility seems not to have been appreciated in capillary adhesion previously. In particular, wetting liquids will  naturally migrate to local minima in the separation between two surfaces because of a geometry-induced capillary pressure gradient \cite{Reyssat2014,Renvoise2009}: whenever a liquid bridge is on a sloped surface, the surface separation is smaller on one side (and hence the suction pressure larger) than the other (see fig.~\ref{fig:DropMigration}). This pressure imbalance causes a pressure gradient, which, in turn, drives the bridge into the gap. (Alternatively, a wetting liquid will minimize its energy by wetting the surfaces more, which, for fixed volume, is achieved by moving to narrower gaps.) This motion has two important effects on the capillary adhesion force: firstly, the liquid bridge is more confined vertically when it reaches the minimum, and so spreads further laterally, increasing the area over which the suction pressure operates. Secondly, the gap separation at the bridge's edge is smaller when at the minimum, and hence the suction pressure is larger (in magnitude). Both of these effects can be seen by comparing the two panels in the lower half of fig.~\ref{fig:schematic}. Since both effects are expected  to increase the force of capillary adhesion on rough surfaces,  they suggest that capillary adhesion may be enhanced on rough surfaces by splitting a single capillary bridge into many smaller bridges, each of which may move to a local minimum of separation between the two surfaces. 

In this paper, we consider how to maximize the adhesion force of bridges in the presence of roughness for a given volume of liquid: is it significantly better to have many smaller bridges rather than one large bridge?  We compare the results derived from a model of many capillary bridges between a rough surface and a plate to the usual simple model of capillary adhesion between two planar surfaces. We also contrast the behaviour on rough surfaces with the detailed study by \citet{DeSouza2008split} of many bridges between parallel, planar surfaces. 
Finally, we note that when there are many capillary bridges beneath a rough surface, the geometry--capillary-induced migration of bridges gives a mechanism to generate a steady viscous shear force; we investigate the magnitude of this force and determine the effect of splitting on this shear force.

\section{Adhesion force when splitting bridges}

\subsection{Simple capillary adhesion} \label{sec:SimpleAdhesion}

A simple model of capillary adhesion is a single, stationary liquid bridge of a given volume $\V$, that is confined between two parallel, rigid plates in the absence of gravity, as shown in the top left panel of fig.~\ref{fig:schematic}. The interfacial tension between the liquid and air, $\gamma$, acts in two distinct ways to provide a force on the plates. Firstly, there is a pressure jump across the free surface of the liquid, given by the Young--Laplace equation
\begin{equation} \label{eq:Young-Laplace}
\Delta p = \gamma \kappa,
\end{equation}
where the local mean curvature of the interface is $\kappa = (1/R_1 + 1/R_2)$, with  $R_1,R_2$ being the principal radii of curvature \cite{deGennes2004}. In equilibrium, the bridge pressure is constant and acts over the wetted area, $A=\pi r^2$, with $r$ the wetted radius. The  pressure force pulling the plates together is then $\FPressure=-\Delta p \times A$. Secondly, the free surface tugs on the plates where the liquid, gas and solid meet; this gives rise to a tension force between the plates of magnitude $\FTension=\gamma \sin \theta \times 2 \pi r$, where $\theta$ is the contact angle (see fig.~\ref{fig:schematic}). The total adhesion force is the sum of these two contributions: $F=\FPressure + \FTension$.

When at rest, a liquid bridge will have a constant mean curvature because there must be no flow, and therefore no pressure gradient within it. (These surfaces of constant curvature, such as catenoids and nodoids, are sometimes referred to as Delaunay surfaces \cite{Delaunay1841}.) To find the capillary adhesion force provided by such a bridge, we may therefore solve for the shape of the bridge and determine the force from this solution.

As a point of comparison, it is helpful to consider the case where the gap width $h$ is much smaller than the other length scales in the system and the bridge wets the plates (small contact angles $\theta$). In this case the capillary bridge radius is much larger than its height and the pressure force is much larger than the tension force, $\FPressure\gg \FTension $. In addition, the meniscus curvature is dominated by the component between the two plates --- the radius of curvature around the bridge is much larger. We can therefore approximate the pressure by $\Delta p \approx -2\gamma \overline{\cos \theta}/h$, where $\overline{\cos \theta}=(\cos\theta_l+\cos\theta_u)/2$ is the average of the cosines of the contact angles $\theta_{l,u}$ on the lower and upper surfaces (if they differ, otherwise $\overline{\cos \theta}=\cos\theta$). We then calculate the dimensionless capillary adhesion force pulling the plates together to be 
\begin{equation} \label{eq:FlatCase}
\frac{\Fflat}{\gamma \V/h^2} \approx 2 \overline{\cos \theta}.
\end{equation}

Crucially, in this approximation, the adhesion force scales linearly with the volume of the bridge $\F \propto \V$. If this is always a good approximation, then it may be expected that there is never a benefit to splitting a given volume of liquid $\Vtot$ into $n$ identical bridges because the total force is independent of the number of bridges: $\Ftot = n \F \propto n \V = \Vtot$. Note that, because of this invariance when splitting, if the volume $\V$ is replaced by $\Vtot$ in \eqref{eq:FlatCase} then $\Fflat$ could instead be considered as the total adhesion force due to $n$ bridges. 


However, as the liquid is divided into smaller bridges, other effects that are neglected when making the approximation \eqref{eq:FlatCase} may become more important --- for example the tension force at the contact line, or the second radius of curvature. In fact, adhesion may be expected to increase with the number of bridges when accounting for the tension force. If each bridge volume can be well approximated by a cylinder, $\V \approx \pi r^2 h$, then the tension force per bridge $\FTension=2\pi r \times \gamma \sin \theta \propto \sqrt{\V}$; the total contribution to the adhesion force by the tension of $n$ bridges of volume $\V=\Vtot/n$ grows like $\sqrt{n}$ for a fixed total volume $\Vtot$. Whether this is a significant enhancement depends on $\theta$, but it may be expected that the total adhesion force increases with splitting if $\theta$ is not too small, as was found in \cite{Cai2008}.

However, simply adding the contributions of the tension force and extra curvature does not accurately portray the behaviour of the adhesion force for a single bridge as the bridge volume decreases. In particular, these approximations do not take account of when stable bridges can no longer be formed and will rupture. It is therefore necessary to make a more detailed calculation of the bridge shape to determine whether dividing into many bridges results in an increase in the adhesion force. 


\subsection{Modelling capillary adhesion} \label{sec:BVPModel}

We consider a liquid bridge confined between two solid surfaces, and neglect the effect of gravity on the bridge (e.g., because the bridge is sufficiently small that the Bond number is negligible).
A liquid bridge in equilibrium cannot include flow and hence must have a uniform internal pressure, given by the Young--Laplace equation \eqref{eq:Young-Laplace}. To find the capillary adhesion force provided by such a bridge, the shape of the bridge must be found first before determining the corresponding force as the sum of Laplace pressure forces and capillary line tension forces.

The exact shape of a single capillary bridge can be determined by solving a set of ordinary differential equations (ODEs) that describe a surface of constant curvature, subject to boundary conditions that impose that the bridge meets the given solid surfaces, does so at contact angle(s) $\theta$ and has a bridge volume $\V$. While this problem shall be solved numerically from ODEs here, note that \citet{Carter1988} gave an analytical solution for a bridge spanning two smooth, parallel plates; here, we approach the problem by solving a series of ODEs because this method does not constrain the geometry to planar surfaces, allowing more versatility to describe other topographies. 

\begin{figure}[tbp]
	\begin{center}
		\includegraphics[width=0.9\linewidth]{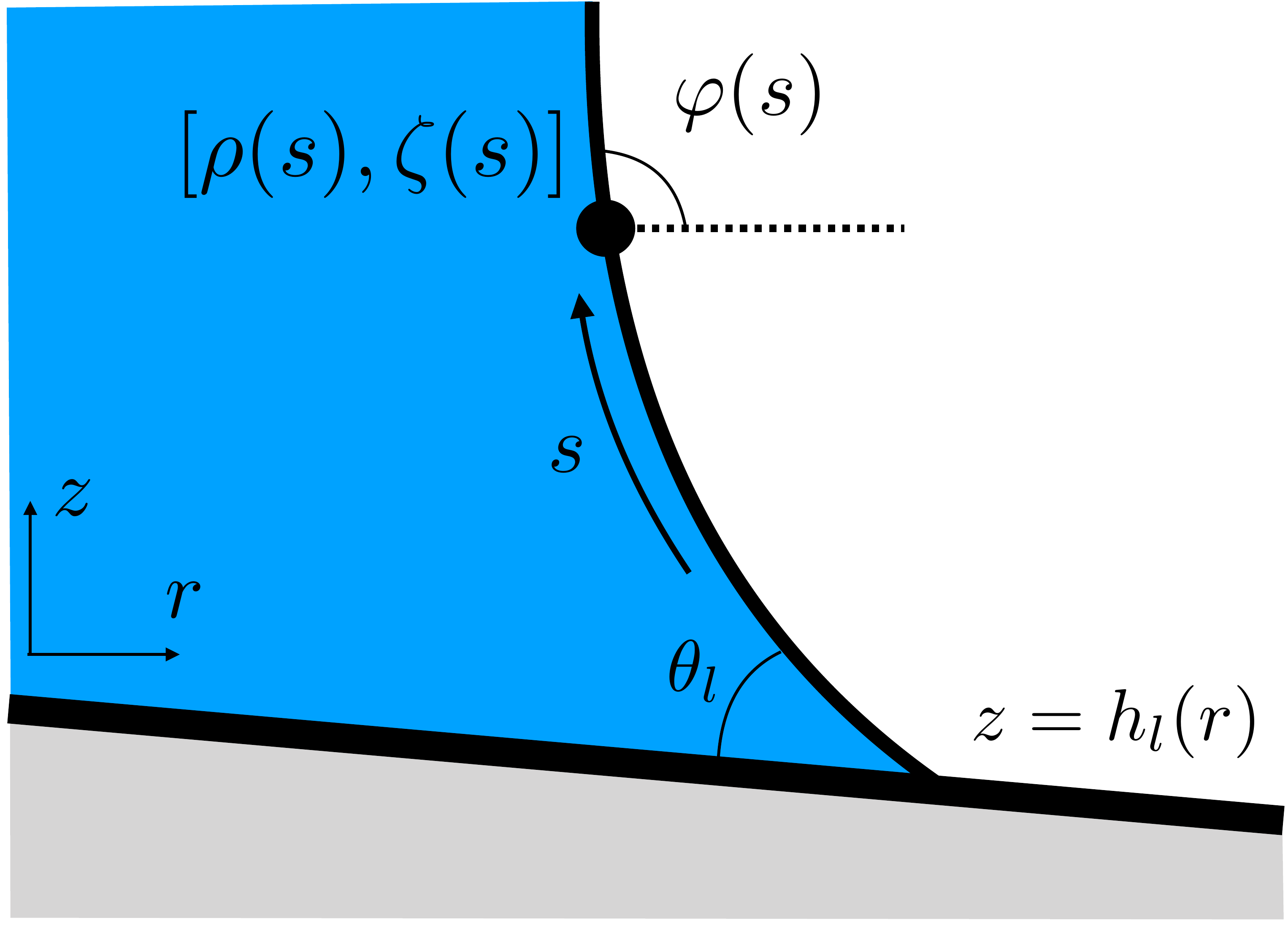} 
		\caption[Definitions for the constant curvature boundary value problem]{The surface of a liquid bridge $(r,z)=[\BVPr(s),\BVPz(s)]$ is parametrized by the arc length, $s$, (measured from one contact line) and is defined in terms of the angle, $\BVPAngle(s)$, between its tangent and the horizontal. (In this diagram, the liquid bridge is to the left of the curve defining the free surface.) The liquid bridge interface meets the lower surface $z=h_l(r)$ at the static contact angle $\theta_l$ (and similarly for the upper surface which is not depicted).} 
		\label{fig:BVPSetup} 
	\end{center}
\end{figure}

The shape of an axisymmetric bridge can be described by a single curve that gives the free surface when it is rotated about the central axis. This curve can be entirely determined by the function $\BVPAngle(s)$, where $\BVPAngle$ is the angle between the tangent to the curve and the horizontal, which is parametrized by the arc length, $s$,  as illustrated in fig.~\ref{fig:BVPSetup}.
The vertical and radial coordinates of this curve $r=\BVPr(s)$ and $z=\BVPz(s)$ are determined from the angle by elementary geometry: $\upd \BVPr/ \upd s = \cos{\BVPAngle}$, $\upd \BVPz / \upd s = \sin{\BVPAngle}$.
The function $\BVPAngle(s)$ for the constant curvature surface satisfies the ODE \citep[see, for example,][]{deGennes2004,Qian2006,Slater2014,Macner2014} 
\begin{equation} \label{eq:ODE_ConstK}
\frac{\mathrm{d} \BVPAngle}{\mathrm{d}s} + \frac{\sin{\BVPAngle}}{\BVPr} = ~\kappa~ = \frac{\Delta p}{\gamma}, 
\end{equation}
where $\kappa$ is the (unknown) constant surface curvature, and the arc length $s\in[0,\smax]$ for some (unknown) total arc length $\smax$. 

This system is subject to boundary conditions that enforce that the bridge meets the given solids at a specific contact angle, $\theta$, as well as an integral constraint imposing the total bridge volume
\begin{multline} \label{eq:BVPVolCon}
\V = \pi \int_0^{\smax} \BVPr(s)^2 \sin \BVPAngle ~\mathrm{d}s \\
- 2 \pi \int_0^{\BVPr(\smax)} x [h_u(\BVPr(\smax))-h_u(x)] ~\mathrm{d}x  \\
- 2 \pi \int_0^{\BVPr(0)} y [h_l(y)-h_l(\BVPr(0))] ~\mathrm{d}y, 
\end{multline}
where $z=h_l(r)$ and $z=h_u(r)$ define the surfaces of the lower and upper solid, respectively. Note that the latter two integrals here account for the volume excluded due to the presence of the solid surfaces within the bridge perimeter. 

This system is simplified by eliminating the unknown constant curvature $\kappa$ by differentiating \eqref{eq:ODE_ConstK} with respect to $s$. For numerical convenience, the integral constraint \eqref{eq:BVPVolCon} is also converted into an additional ODE by introducing a function for the cumulative volume $\int_0^{s} \pi \BVPr (\eta)^2 \sin \BVPAngle (\eta) ~\mathrm{d}\eta$. For given solid surfaces separated by a typical lengthscale $L$ (that shall be defined concretely for each specific problem), the problem is non-dimensionalized by scaling all extrinsic lengths by $L$, while the meniscus arc length is scaled by the total arc length, $\smax$. These different lengthscales introduce a dimensionless parameter $\Smax=\smax/L$. The problem of finding the shape of a constant curvature liquid bridge then becomes that of solving 
\begin{equation}
\begin{aligned} 
\frac{\upd^2\BVPAngle}{\upd S^2} &+ \Smax\frac{\cos{\BVPAngle}}{\BVPR} \frac{\upd\BVPAngle}{\upd S} - \Smax^2 \frac{\sin{\BVPAngle} \cos{\BVPAngle}}{\BVPR^2} = 0, \label{eq:BVP}\\
&\frac{\upd \BVPR}{\upd S} = \Smax\cos{\BVPAngle},  \qquad  
\frac{\upd \BVPZ}{\upd S} = \Smax\sin{\BVPAngle}, \\
&\frac{\upd v}{\upd S} = \Smax\pi \BVPR^2 \sin{\BVPAngle},  \qquad
\frac{\upd  \Smax}{\upd S} = 0, 
\end{aligned}
\end{equation}
where $S = s/\smax \in [0,1]$ is the (scaled) arc length; $\BVPR(S) = \BVPr(s)/L$ and $\BVPZ(S)=\BVPz(s)/L$ are the dimensionless radial and vertical coordinates of the meniscus; $\BVPAngle$ is the inclination angle as defined previously; $v(S)=\int_0^{S} \Smax \pi \BVPR(x)^2 \sin [\BVPAngle(x)] ~\mathrm{d}x$ is the dimensionless cumulative volume and $\Smax=\smax/L$ is the dimensionless total arc length. The positions of the solid surfaces are now given by  $\tilde{z}=H_l(\tilde{r}),H_u(\tilde{r})$ where $H_{u,l}=h_{u,l}/L$ and the dimensionless cylindrical coordinates are $(\tilde{r},\tilde{z})=(r/L,z/L)$.

This is a sixth order system of ODEs and so six boundary conditions are required. Two of the required boundary conditions are simply that the bridge interface meets each solid surface
\begin{equation} \label{eq:BVPBCstart}
\BVPZ(0) = H_l[\BVPR(0)], \qquad 
\BVPZ(1)=H_u[\BVPR(1)]. 
\end{equation}

Two further conditions are given by the contact angles at each end, carefully accounting for the slope of each of the solids
\begin{equation} \label{eq:BVPBC_CA}
\begin{aligned}
\BVPAngle(0) =& \pi-\theta_l + \tan^{-1} \left\{ H_l'[\BVPR(0) ] \right\}, \\
\BVPAngle (1) =& \theta_u + \tan^{-1} \left\{ H_u'[\BVPR(1)] \right\},
\end{aligned}
\end{equation}
where the dash notation denotes differentiation with respect to $\tilde{r}$, and $\theta_l$ and $\theta_u$ are the contact angles at each of the solids.

The final two boundary conditions enforce the fixed volume condition \eqref{eq:BVPVolCon}
\begin{equation} \label{eq:BVPBCend}
\begin{aligned}
&v(0) = 0, \\
v(1) = v_1 + 2 \pi &\int_0^{\BVPR(1)} x [H_u(\BVPR(1))-H_u(x)] ~\mathrm{d}x \\
+ 2 \pi &\int_0^{\BVPR(0)} y [H_l(y)-H_l(\BVPR(0))] ~\mathrm{d}y, 
\end{aligned}
\end{equation}
where $v_1=\V/L^3$.

In principle, solving the coupled ODEs \eqref{eq:BVP} subject to the boundary conditions \eqref{eq:BVPBCstart}--\eqref{eq:BVPBCend} gives the shape of the bridge $(\BVPR,\BVPAngle, \BVPZ)$ for given boundaries $H_l, H_u$, contact angles $\theta_l,\theta_u$ and bridge volume $v_1$. 

Once the shape of the bridge is known, the force $\mathbf{F}_1$ applied by this bridge on one of the solid surfaces can be calculated as the sum of the contributions from the capillary pressure and tension terms 
\begin{equation}
\mathbf{F}_1 = \int_D - \Delta p ~\mathbf{n} ~\upd A + \int_{\partial D} \gamma \mathbf{t} ~\upd s,
\end{equation}
where $\mathbf{n}$ is the normal to the solid surface, $\mathbf{t}$ is the tangent to the liquid interface, $D$ is the wetted area and $\partial D$ its perimeter. Note that the capillary pressure jump, $\Delta p$, is determined from the shape of the bridge through the (constant) meniscus curvature, $\kappa$, due to the Young--Laplace equation \eqref{eq:Young-Laplace}, $\Delta p = \gamma \kappa$; the dimensionless curvature $K = \kappa L$ is given by 
\begin{equation} \label{eq:ConstCurvature}
K = \frac{1}{\Smax}\frac{\mathrm{d} \BVPAngle}{\mathrm{d} S} + \frac{\sin \BVPAngle}{\BVPR}.
\end{equation}

After finding the adhesion force due to the surface tension of one bridge, $\F$, the total adhesion force, $\Ftot=n\F$, for $n$ identical bridges with a fixed total volume $\Vtot=n\V$ can be calculated.

\subsection{Bridge splitting on flat surfaces} \label{sec:FlatSplit}

Let us return to the problem of splitting bridges between two parallel plates, as shown in the left side of fig.~\ref{fig:schematic}. We would like to know whether having many bridges gives stronger adhesion and, if so, how to maximize the adhesion force by dividing the liquid into many bridges. This problem has been previously studied by \citet{DeSouza2008split} using an energy minimization approach. Here the work of \citet{DeSouza2008split} is reproduced by solving the set of ODEs above; we shall move to consider the effect of surface roughness in \S\ref{sec:RoughSplitting}. 

Consider two identical parallel, rigid plates that are a fixed distance $h$ apart. A volume $\Vtot$ of liquid of surface tension $\gamma$ is divided into $n$ identical bridges of equal volume $\V=\Vtot/n$ that meet each plate at a contact angle $\theta$ (for simplicity, this is taken to be the same on each surface). This scenario is illustrated in the bottom left panel of fig.~\ref{fig:schematic}. The total adhesion force, $\Ftot$, due to these $n$ bridges is found by solving the ODEs  \eqref{eq:BVP} for the shape of a single bridge of volume $\V$, subject to the boundary conditions \eqref{eq:BVPBCstart}--\eqref{eq:BVPBCend} with flat surfaces $H_l=0$, $H_u=1$ (using the lengthscale $L=h$). From the bridge shape, the adhesion force from this single bridge, $\F$, is calculated as the sum of the contributions from the suction pressure within the bridge and the tension force around the bridge meniscus, $F=\FPressure+\FTension=-\pi \gamma r^2 \kappa + 2\pi\gamma r \sin \theta$, where the radius $r$ is evaluated at one of the solid surfaces ($z=0,h$); the total adhesion force is then $\Ftot=n \F$. 

\begin{figure}[tbp]
	\centering
	\includegraphics[width=\linewidth]{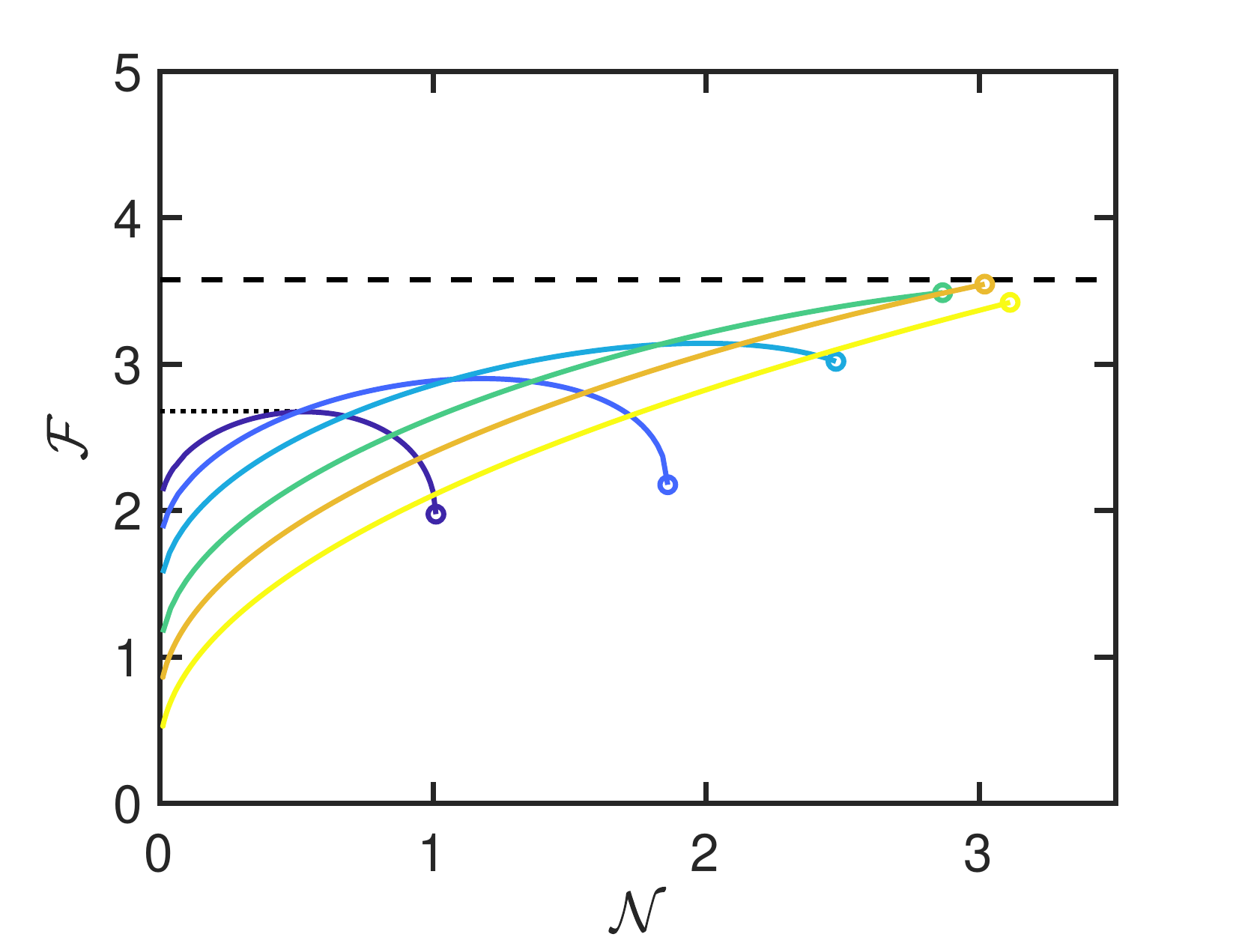}
	\caption[Total adhesion force when splitting bridges between parallel plates]{The dimensionless total capillary adhesion force, $\Fbridge = \Ftot/(\gamma\Vtot/h^2)$, as a function of the scaled bridge number, $\nbridge=n/(\Vtot/h^3)$, obtained when a liquid volume $\Vtot$ with surface tension $\gamma$ is split into $n$ bridges that are confined between two parallel plates, separated by a distance $h$. Results are shown for various different values of the contact angle: $\theta=0^\circ, 30^\circ, 45^\circ, 60^\circ, 70^\circ, 80^\circ$ (coloured dark blue to yellow). Open circles denote the furthest level of bridge splitting possible before the bridges can no longer exist in equilibrium and must rupture. The dashed line denotes the maximum adhesion force $\Fbridge=3.58$, as found by \citet{DeSouza2008split}; the dotted line is the maximum adhesion force when $\theta=0$, $\Fbridge=2.68$.}
	\label{fig:FlatForce}
\end{figure}

Results for the adhesion force when splitting bridges between parallel plates are presented in dimensionless form in fig.~\ref{fig:FlatForce}; when comparing to real data, we must fix the parameter $\Vtot/h^3$ and then consider discrete points along each curve, corresponding to integer values of $n$. The results of fig.~\ref{fig:FlatForce} agree quantitatively with those presented by \citet{DeSouza2008split} or \citet{Carter1988}, confirming the accuracy of the numerical approach. These benchmark results will also be useful for comparison with the results for rough surfaces later. 

Figure \ref{fig:FlatForce} shows the effect of bridge splitting on the adhesion force obtained for bridges between two identical flat surfaces with contact angles $\theta=0, 30^\circ, 45^\circ, 60^\circ, 70^\circ$ and $80^\circ$. For small contact angles, a maximum adhesion force is attained at an intermediate number of bridges, well before the solutions cease to exist and the bridges must break. With larger contact angles, however, the maximum adhesion force is attained when the bridges are only just able to stably bridge the gap: any further splitting would result in breaking of the bridges and no adhesion. The point at which there ceases to be a stable solution shall be referred to as `rupture', though we emphasise that this is purely a static definition of the point by which rupture must have occurred.

In general, as the fixed liquid volume is divided into more bridges, the (positive) azimuthal curvature around each bridge increases, which will act to reduce the suction pressure force $\FPressure$; meanwhile, the total contact perimeter increases, resulting in an increase in the tension force $\FTension$. The enhanced importance of the tension force leads to the increase in adhesion force as the bridges are further split for larger contact angles; for highly wetting liquids (smaller $\theta$) the tension contribution is smaller, and eventually is less important than the curvature changes. (This is not quite true for perfectly wetting liquids, $\theta=0$, which have no tension force --- see Appendix \ref{app:PerfectWetting} for an overview of this case.) 

\citet{DeSouza2008split} found that, over all contact angles $\theta$ and number of bridges $n$, the maximum adhesion force is $\Ftot \approx 3.58 \times (\gamma\Vtot/h^2)$ which is obtained when $\theta \approx 70^\circ$ and the bridges are just at the point of rupture (i.e.~there are no solutions with smaller bridge size). For perfectly wetting liquids, $\theta=0$, the increase due to splitting is approximately one third larger than expected on the basis of the simple estimate given by eqn.~\eqref{eq:FlatCase}: the numerical solutions suggest that a maximum force $\Ftot \approx 2.68 \times (\gamma\Vtot/h^2)$ does exist, so that there is still a small benefit to bridge splitting, even in the perfectly wetting case where there is no contribution from the capillary tension force. Across all contact angles studied, it is found that dividing the liquid into many bridges only leads to a mild improvement in the adhesion force (i.e.~not orders of magnitude larger). A key question is then: is the enhancement from bridge splitting more significant with a rough surface?

\subsection{Bridge splitting on a rough surface} \label{sec:RoughSplitting}

Capillary adhesion between rough surfaces has previously been considered by \citet{Cai2008}, although they assume a circular shape for each bridge meniscus --- an assumption that falls down when the bridge is no longer very wide compared to its height (as will eventually be the case when splitting bridges). Similar to the flat case, we shall find that the details of the bridge shape and when rupture occurs are crucial in determining whether splitting has a beneficial effect. We therefore turn now to investigate the bridge shape and adhesion force in this rough scenario in some detail.

\subsubsection*{Problem set-up}

A given volume of liquid, $\Vtot$, is split into $n$ identical bridges, each of volume $\V=\Vtot/n$. As is the case throughout this work, the liquid has a surface tension $\gamma$ and makes a contact angle $\theta$ with each of the solid surfaces. The bridges are sandwiched between two rigid surfaces, and are expected to be located at local minima in the separation between the surfaces (because of geometry-induced capillary pressure gradients, as illustrated in fig.~\ref{fig:DropMigration}). For simplicity, one surface is considered to be planar and the other rough; the most important detail is how the distance between the two surfaces varies, which cannot distinguish between two rough surfaces and a rough-smooth pair. (However, the geometry osf the surfaces does matter somewhat; the contact angle conditions at the triple contact line \eqref{eq:BVPBC_CA} differ between the rough-rough and rough-planar cases.)

\begin{figure}[tbp]
	\centering
	\includegraphics[width=\linewidth]{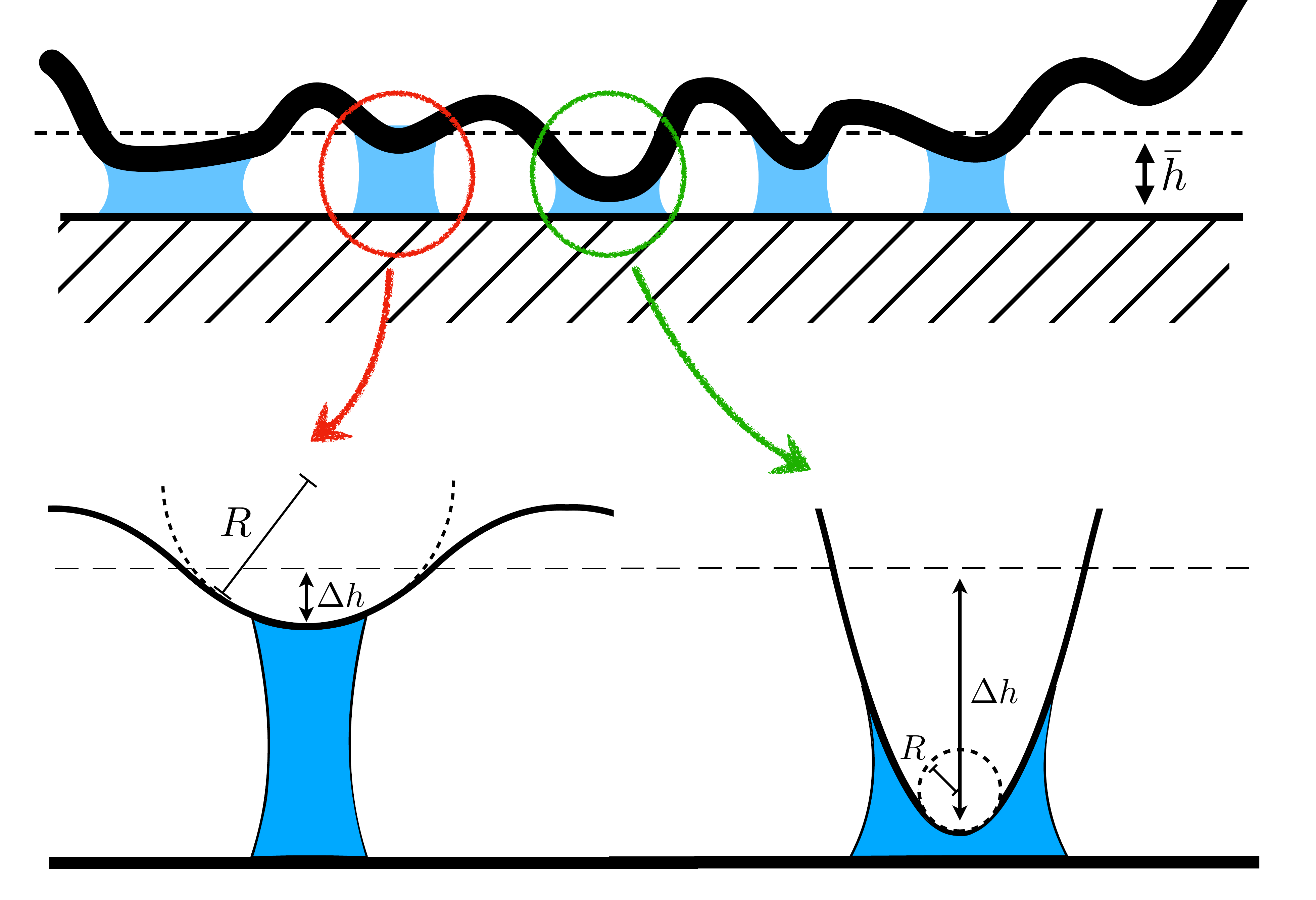}
	\caption[Illustration of different roughnesses]{A schematic showing the roughness parameters for two different asperities that are approximated locally as parabolas. The left asperity has $\Dh/\h = 0.2 $, whilst the right one has $\Dh/\h=0.9$. Both asperities shown have the same aspect ratio $\asp$ defined in eqn.~\eqref{eq:AspectRatio}. (Bridge shape and volumes are not to scale.) }
	\label{fig:VariousRoughness}
\end{figure}

We now need to choose a model for the surface roughness. Consider a rough surface separated from a plane by an average distance $\h$, as shown in fig.~\ref{fig:VariousRoughness}. Assuming that a single asperity is locally smooth and axisymmetric, then, since it is a local minimum of separation, the shape of the asperity close to its tip can be approximated by a parabola: $h(r)=\ho+r^2/2\R$, for some $\ho<\h$ and $\R$, with $r$ the radial distance from the centre. (Analogous calculations with linear cone-like asperities are given in Appendix \ref{app:Variations} for comparison.) We therefore assume that each of the $n$ bridges is confined between a plane and one of these parabolas. While surface roughness is expected to have a variety of roughness amplitudes and wavelengths, as shown in fig.~\ref{fig:VariousRoughness}, we assume each asperity has the same curvature and amplitude. Surfaces with the same mean separation, $\h$, are compared by varying the amplitude of the roughness, ${\Dh=\h-\ho}$, but fixing the aspect ratio 
\begin{equation} \label{eq:AspectRatio}
\asp=\frac{\R\Dh}{\h^2}.
\end{equation} 
The aspect ratio $\asp$ gives the relative size of the width of the roughness $\sqrt{2 \R \Dh}$ to its average height $\h$. Examples of two asperities with the same aspect ratio $\asp$ and different roughness amplitudes $\Dh$ are shown in the lower half of fig.~\ref{fig:VariousRoughness} (for an example of different aspect ratios, see the inset to fig.~\ref{fig:Asp_Sawtooth} in Appendix \ref{app:Variations}). 

For given parameters, the boundary value problem (BVP) defined in \S\ref{sec:BVPModel} is solved using a non-dimensionalization with the lengthscale $L=\h$. More specifically, the ODEs \eqref{eq:BVP} are solved for the shape of a single bridge of volume $\V$, subject to the boundary conditions \eqref{eq:BVPBCstart}--\eqref{eq:BVPBCend} when one surface is flat, $H_l=0$, and the other has parabolic shape, $H_u(\tilde{r})=\ho/\h + \tilde{r}^2/2(\R/\h)$. 
These boundary conditions can be written as
\begin{equation} \label{eq:BCs}
\begin{aligned}
\BVPZ(0) = 0, \quad
&\BVPZ(1) =  \frac{\ho}{\h} + \frac{[\BVPR(1)]^2}{2(\R/\h)}, \\
\BVPAngle (0) = \pi - \theta, \quad
&\BVPAngle (1) = \theta + \tan^{-1} \left[ \frac{\BVPR(1)}{(\R/\h)} \right],  \\
v(0) = 0, \quad
&v(1) = \frac{\V}{\h^3} + \frac{\pi}{4} \frac{[\BVPR(1)]^4}{(\R/\h)}.
\end{aligned}
\end{equation}
Once the aspect ratio ($\asp = \R \Dh/\h^2$), the dimensionless roughness amplitude ($\Dh/\h$), and the dimensionless bridge volume ($\V/\h^3$), are specified, then the values of $\ho/\h$ and $\R/\h$ to be used in these boundary condition are determined by the relations
\begin{align}
\frac{\ho}{\h} &= 1-\frac{\Dh}{\h}, \\
\frac{\R}{\h} &= \frac{\asp}{\Dh/\h}.
\end{align}

For each choice of parameters, the system of ODEs \eqref{eq:BVP} is solved subject to the boundary conditions \eqref{eq:BCs} using MATLAB's in-built BVP solver \texttt{bvp4c}. The adhesion force provided by a single bridge on the planar surface can then be calculated by 
\begin{equation}
\frac{\F}{\gamma \V/\h^2} =  \frac{-\pi K \BVPR(0)^2 + 2 \pi \BVPR(0) \sin \theta}{\V/\h^3},
\end{equation}
where the dimensionless curvature $K$ is defined in \eqref{eq:ConstCurvature}. (Recall that a positive adhesion force corresponds to the plate being pulled towards the rough surface --- adhesion.)

In fact, the dimensionless adhesion force for a single bridge is precisely the dimensionless total adhesion force, $\Ftot=n\F$, from $n$ such bridges with combined volume $\Vtot=n\V$, which we write as
\begin{equation}
\Fbridge \equiv \frac{\Ftot}{\gamma \Vtot/\h^2}=\frac{\F}{\gamma \V/\h^2}.
\end{equation}
This dimensionless force can be expressed as a function of the scaled number of bridges, which we define by
\begin{equation}
\nbridge \equiv \frac{\h^3}{\Vtot} n = \left( \frac{\V}{\h^3} \right)^{-1}.
\end{equation}

If the two surfaces are sufficiently separated from one another, or equivalently if the bridge volume is sufficiently small, then a stable liquid bridge cannot be formed between the two solid surfaces. Particular care must be taken to calculate when this loss of equilibrium, which we associate with bridge rupture, occurs. The scaled bridge number, $\nbridge$, at which this occurs (as a function of $\Dh$) is calculated using \texttt{AUTO-07P},  a computational tool for finding bifurcations by arc-length continuation of solutions in multiple parameters \citep{Doedel2007}. Starting from a numerical solution to the BVP --- given by the ODEs \eqref{eq:BVP} and boundary conditions \eqref{eq:BCs} --- for the flat case, $\Dh=0$, the scaled number of bridges $\nbridge$ is increased until a bifurcation is detected before following this bifurcation as $\Dh$ is increased. This gives the locus of the rupture bifurcation $\nbridge_{\text{rup}}(\Dh)$. 

Attention is restricted only to those solutions where the bridge is shorter than the average roughness, $\xi(1)<1$; beyond this limit, the bridges take up a significant portion of each roughness asperity and so approximating the shape as a parabola may no longer be applicable.
Additionally, it may reasonably be expected that optimal adhesion through splitting should be achieved when bridges are relatively close to rupture (as was found for flat plates in \S\ref{sec:FlatSplit}). Therefore, the adhesion force is calculated as a function of the roughness amplitude $\Dh$ and the number of bridges $\nbridge$ between the upper limit on $\nbridge$ where rupture occurs and the lower limit where the meniscus height is equal to the mean roughness height, $\xi(1)=1$.

\subsubsection*{Results for the adhesion force}


In the results presented here, we focus on the case of liquid that perfectly wets both surfaces, $\theta=0$, and an aspect ratio $\asp=10$. In Appendix \ref{app:Variations} it is shown that the effects of varying the contact angle and aspect ratio are largely quantitative rather than qualitative with a relatively minor impact on the results (this is particularly the case for the effect of the aspect ratio $\asp$). 

\begin{figure}[tbp]
	\centering
	\includegraphics[width=0.95\linewidth]{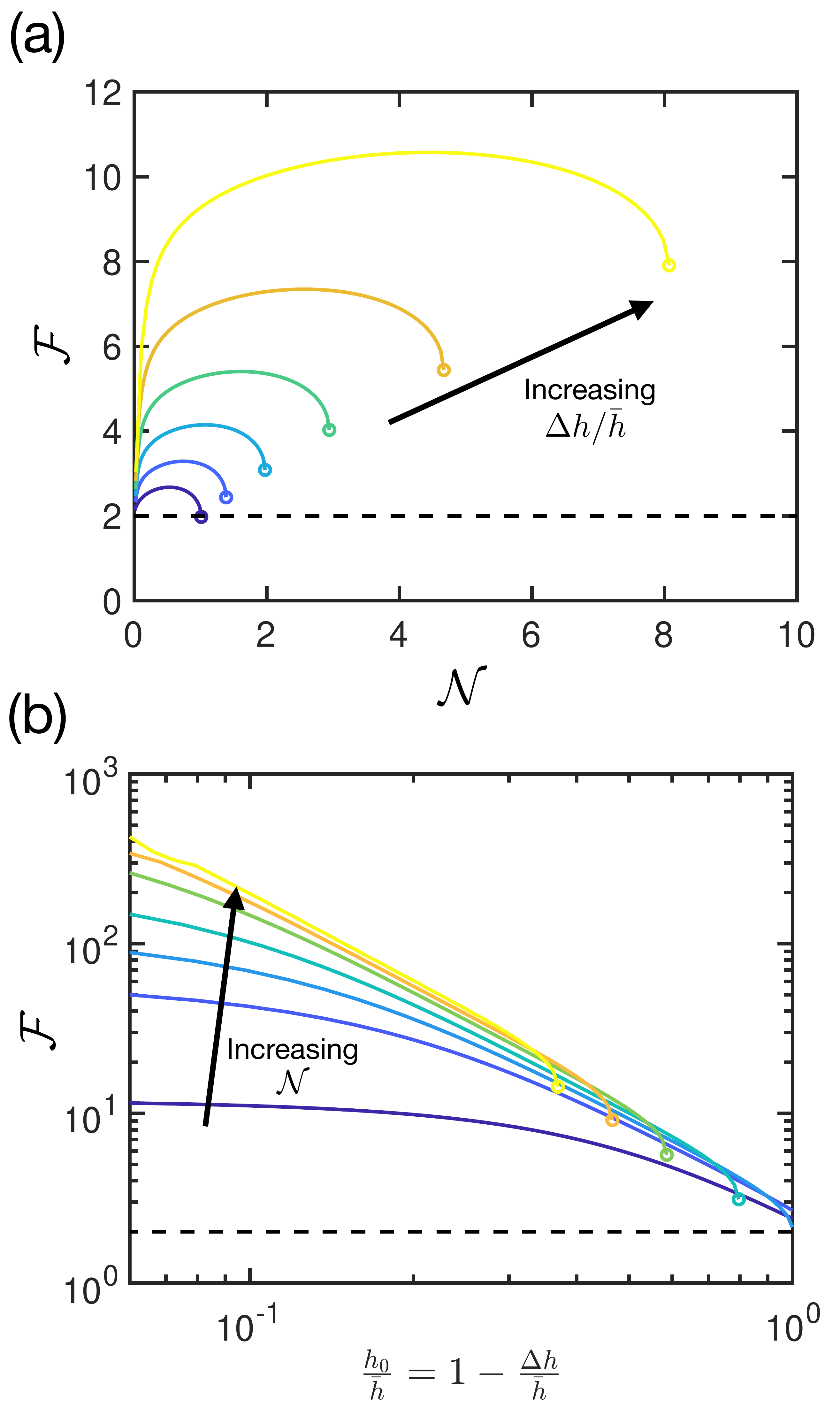}
	\caption[Total adhesion force when splitting bridges on a rough surface]{The dimensionless total adhesion force as a function of (a) the scaled number of bridges, $\nbridge=n/(\Vtot/\h^3)$, at fixed roughness amplitudes $\Dh/\h=\{0,0.1,0.2,0.3,0.4,0.5\}$ (plotted up to rupture, which is indicated by circles), and (b) the minimum gap width, $\ho/\h=1-\Dh/\h$, with fixed number of bridges $\nbridge =\{0.1,0.5,1,2,5,10,20\}$. In both cases here $\theta = 0$ and $\asp = 10$, and results are compared to a dashed line denoting eqn.~\eqref{eq:FlatCase}. Note that in (a), the case $\Dh=0$ corresponds to the case of flat plates studied by \citet{DeSouza2008split}.}
	\label{fig:RoughForce}
\end{figure}

The total adhesion force, $\Ftot$, from $n$ bridges at several values of the fixed roughness amplitude, $\Dh/\h$, is shown in dimensionless form in fig.~\ref{fig:RoughForce}a. (As in the flat case, when comparing these results to real data, one should consider discrete points along the curve, corresponding to integer values of $n$ --- a continuous curve is presented here to account for all possible values of $\nbridge$.) It can be seen that (provided that splitting is actually feasible, i.e.~the total liquid volume $\Vtot$ is not too small compared to $\h^3$) splitting can result in a significant increase in adhesion force. The maximum adhesion force at fixed $\Dh/\h$ can be several times larger than the respective simplified flat case, given in eqn.~\eqref{eq:FlatCase}; moreover, the maximum occurs far from the end-point of splitting, where bridges rupture --- for a given roughness amplitude, there is an optimal level of bridge splitting, achieved without those bridges rupturing. Finally, the curves of $\Fbridge$ as a function of $\nbridge$ are relatively wide and flat, meaning that even levels of splitting that are far from optimal are still able to achieve close-to-optimal adhesion forces.  

Figure \ref{fig:RoughForce}a also suggests that the total adhesion force increases with increasing roughness, and so it is important to also consider varying the roughness amplitude $\Dh/\h$ at fixed number of bridges $\nbridge$; the results are therefore shown in fig.~\ref{fig:RoughForce}b as a function of the minimum gap width $\ho=\h-\Dh$. Here, it is apparent that the adhesion force increases as the roughness amplitude increases ($\ho$ decreases). In most of the cases shown, the force increases by several orders of magnitude as the roughness is increased.

\begin{figure}[tbp]
	\centering
	\includegraphics[width=0.9\linewidth]{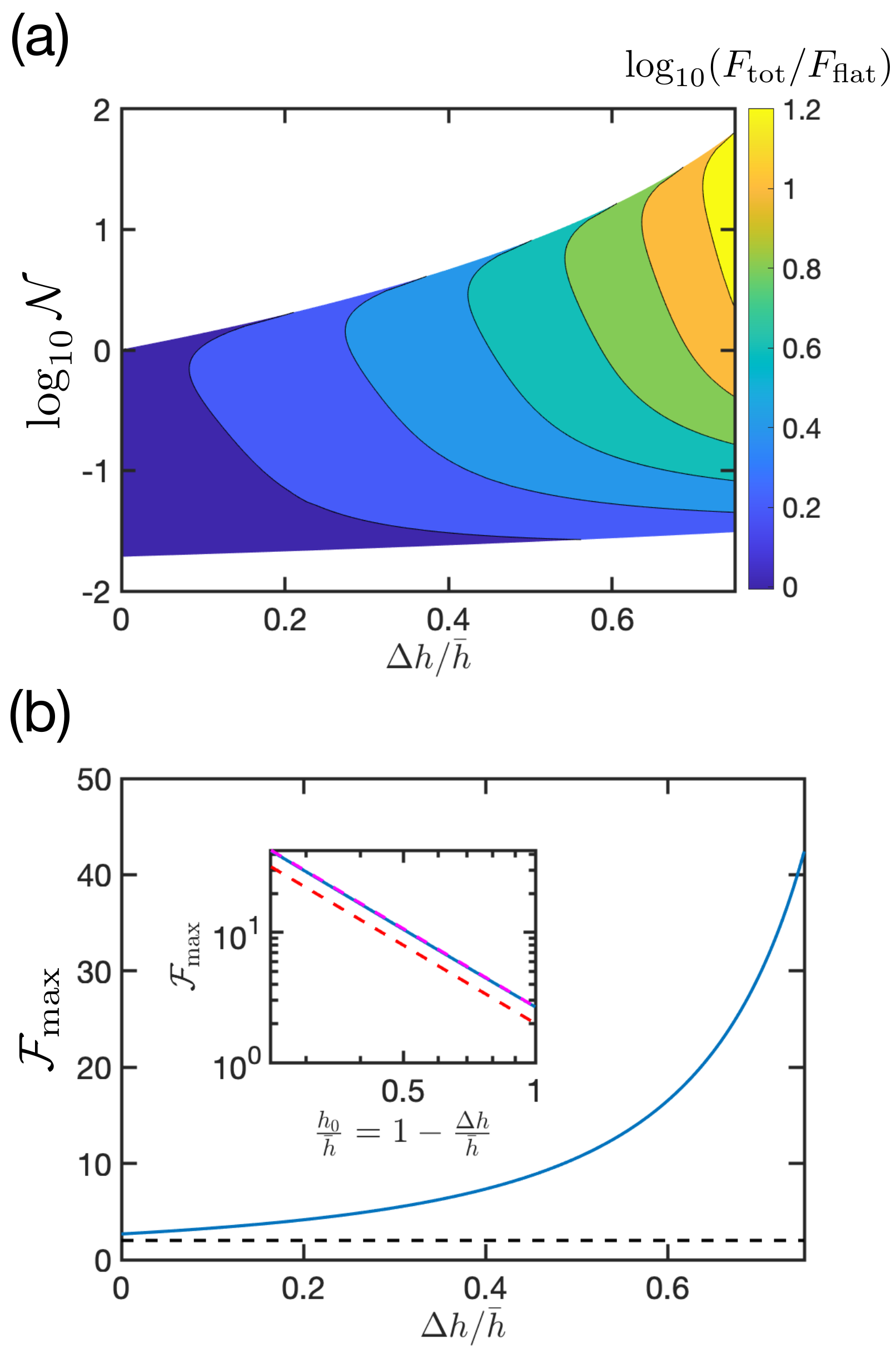}
	\caption[Total adhesion force colour map and maximum adhesion force when splitting]{(a) The relative increase in the total adhesion force, $\Ftot$, for $n$ bridges under a rough surface compared to the simplified force calculation for a single bridge between flat surfaces, $\Fflat$ from eqn.~\eqref{eq:FlatCase}. Here $\Ftot/\Fflat$ is represented by colour (see colour bar), while $\Dh/\h$ represents the amplitude of the roughness and  $\nbridge=n/(\Vtot/\h^3)$ the (scaled) number of bridges. No solutions exist in the upper white region (rupture), whilst solutions where the bridge spreads further than one asperity, $\xi(1)>1$, are ignored (lower white region). 
	(b) The maximum total adhesion force, $\Fmax$, at each roughness amplitude $\Dh/\h$ (solid blue curve) compared to the benchmark scaling of eqn.~\eqref{eq:FlatCase}, shown by the black dashed line. 
	Inset: a log-log plot showing the maximum force (solid curve) as a function of $h_0 = \h - \Dh$. The red (lower) dashed line in the inset denotes $y=2/x^2$, whilst the magenta (upper) dashed line is $y=2.68/x^2$; these respectively correspond to the simple relation \eqref{eq:FlatCase} and to optimal adhesion, both when splitting between flat plates with a separation $\ho$.
	In both (a) and (b), the contact angle and aspect ratio are $\theta=0$ and $\asp=10$, respectively.}
	\label{fig:RoughDropSplit}
\end{figure}

The results for the adhesion force as a function of the splitting and roughness are summarized in fig.~\ref{fig:RoughDropSplit}a. The colour denotes the (logarithm of the) ratio of the force obtained with a particular surface roughness to that obtained in the simplified flat case described by eqn.~\eqref{eq:FlatCase}. From this it can clearly be seen that the force generally increases with splitting (although with an intermediate maximum) and with roughness. Importantly, it is possible to get an increase in the adhesion force of over an order of magnitude compared to the flat case.

Plotting the maximum force achieved by splitting as a function of the roughness amplitude (i.e.~the envelope of curves in fig.~\ref{fig:RoughForce}b), fig.~\ref{fig:RoughDropSplit}b is obtained. The maximum force increases with the relative roughness of the surface, appearing to diverge as $\ho$ decreases like ${\Fmax\sim\ho^{-2}}$ (recall that $\ho$ is the minimal gap width). Indeed, the maximum force is greater than that obtained by the simple flat scaling of eqn.~\eqref{eq:FlatCase} with  $h=h_0$, but is actually less than the maximum force possible if $n$ bridges adhere flat plates separated by $\ho$, i.e.~$\Fmax<2.68 \times (\gamma \Vtot/\ho^2)$ (see fig.~\ref{fig:FlatForce}); however, note that the difference is almost indistinguishable at the scale of the inset to fig.~\ref{fig:RoughDropSplit}b. 


We have seen that bridge splitting may give a large increase in the capillary adhesion force between rough surfaces. The mechanism for this enhanced effect was the migration of bridges to the local minima in the surface separation, which simultaneously increases the capillary suction pressure and the wetted area over which this suction pressure acts. However, this migration of liquid bridges also provides a mechanism through which tangential motion between two rough surfaces can be resisted ---  a capillary-induced resistance to shear. We therefore turn now to quantify this.

\section{Resistance to shear} \label{sec:ShearSplitting}


In this section, we shall explore a capillary-based mechanism for generating resistance to shear. 
This is motivated by experiments on ants which showed a resistance to shear that increases linearly with sliding velocity \citep{Federle2004}. This linear relationship suggests a hydrodynamic mechanism, but to our knowledge has not been explained previously. Here, we investigate whether geometry-induced migration can provide a sufficient shear resistance: we consider a liquid bridge trapped beneath a parabolic asperity that is dragged by an imposed shear away from the minimum below a particular asperity (as in fig.~\ref{fig:DropMigration}, if the planar base is sheared to the right at a speed $U$). The resulting geometric asymmetry introduces a pressure difference that leads to an opposing flow within the liquid bridge. Ultimately, the bridge may be expected to reach a steady state in which this capillary-driven flow precisely balances the dragging caused by the imposed shear. The bridge thus remains beneath the asperity (without being smeared out); at the same time, the capillary-induced back-flow applies a force that resists the imposed shear.

\subsection{Model of the shearing motion}

Consider a short, but wide wetting liquid bridge confined between a flat plate and a rough surface with a parabolic profile $h(r)=h_0+r^2/2\R$ (as in \S\ref{sec:RoughSplitting}). The flat plate is sheared at a constant speed $U$ in the positive $x$-direction, with the fluid flow modelled by thin-film lubrication equations \citep[see Chapter 5 of][]{Leal2007}. The horizontal velocity in the liquid is then
\begin{equation}
\mathbf{u} = -\frac{\nabla p}{2\mu}z(h-z)  + \mathbf{U}\left(1-\frac{z}{h} \right),
\end{equation}
where $\nabla = (\partial_{x},\partial_{y})$ is the gradient operator in the horizontal plane, $\mathbf{U}=(U,0)$ is the applied shear velocity in the $x$-direction and $\mu$ is the dynamic viscosity of the liquid. Integrating this in $z$ gives the depth-integrated volumetric flux
\begin{equation} \label{eq:VolFlux}
\mathbf{q} = -\frac{h^3}{12\mu} \nabla p + \frac{h}{2}\mathbf{U}. 
\end{equation} 

Focusing on steady states, for simplicity, conservation of mass \citep{Leal2007} then shows that the volumetric flux $\mathbf{q}$ must obey the steady-state Reynolds' equation: 
\begin{equation} \label{eq:SteadyMassCons}
\nabla \cdot \mathbf{q}=0 \quad \implies \quad U \frac{\partial h}{\partial x} = \frac{1}{6\mu} \nabla \cdot (h^3 \nabla p),
\end{equation} 
which is reminiscent of the governing equations for coating problems (see, for example, \cite{Levich1942,Wilson1982}).

In equilibrium with no shear, $U=0$, the bridge has its meniscus at $r=\Reql$, and its volume is given by $V=2\pi\int_0^{\Reql}r(h_0+r^2/2\R)~\upd r = \pi (\ho \Reql^2 + \Reql^4/4\R)$, where $(r,\phi)$ are the usual polar coordinates taken in the plane of the flat surface. Note that when the bridge volume is small, $V/(\ho^2 R) \ll 1$, then the bridge width is small compared to the curvature and the bridge experiences a relatively flat surface with approximately constant gap width, so that $V\approx \pi \ho \Reql^2$.

At the meniscus, $r=r_M(\phi)$, the action of surface tension causes a jump in the normal stress proportional to the curvature of the interface. When using the lubrication approximation this simplifies to a jump in the fluid pressure only (since the dominant fluid stress is due to the capillary pressure), which satisfies the Young--Laplace equation locally 
\begin{equation} \label{eq:SplittingYL}
 p (r_M) =  -\frac{2 \gbar}{h(r_M)},
\end{equation}
where $\gbar=\gamma \overline{\cos \theta}$, the external atmospheric pressure is taken as the pressure datum, $p=0$, and the curvature has been approximated by a circular arc between the surfaces. 

\subsubsection*{Non-dimensionalization}

The problem is non-dimensionalized using the height scale $[h]=\ho$, the radial scale $[r]=\sqrt{R\ho}$ and the pressure scale $[p]=\gbar/\ho$. The new dimensionless coordinates are written with hats, so that, for example, the dimensionless separation of the two surfaces is now $\hdimless(\rdimless)=1+\rdimless^2/2$.

The dimensionless version of \eqref{eq:SteadyMassCons} enforces that the dimensionless liquid pressure $\pdimless (\rdimless,\phi)$ 
must obey the partial differential equation (PDE)
\begin{equation} \label{eq:ShearPDE}
\Ca \frac{\partial \hdimless}{\partial \xdimless} = \frac{1}{6} \nabla \cdot (\hdimless^3 \nabla \pdimless),
\end{equation}
where the (modified) capillary number 
\begin{equation} \label{eq:Ca}
\Ca = \frac{\mu U}{\gbar} \sqrt{\frac{\R}{\ho}},
\end{equation}
measures the relative size of the shear (Couette) contribution to the liquid flux compared to the capillary pressure-driven (Poiseuille) flux. 
Note that the definition of $\Ca$ in \eqref{eq:Ca} differs from the usual capillary number \citep{Leal2007}, $\mu U/\gamma$, since the typical vertical and radial length scales are different here: $\ho$ for Couette flow and $\sqrt{\ho \R}$ for the capillary-driven Poiseuille flow. The system is then controlled by two parameters, the capillary number $\Ca$ and the dimensionless bridge volume $\Vdimless$.
We now assume that $\Ca \ll 1$ 
allowing us to treat the shear as a small perturbation to the bridge equilibrium. 

\subsubsection*{Small $\Ca$ perturbation expansion}

To find solutions in the small shear limit, $\Ca\ll1$, a perturbation expansion is conducted in $\Ca$ about the equilibrium solution to determine the pressure field and bridge position. In particular, the position of the meniscus, $\rdimless_M(\phi)$, and the pressure, $\pdimless$, are expanded in terms of the capillary number 
\begin{equation}
\begin{aligned}
\rdimless_M(\phi) &= \REQL + \Ca \, f(\phi) + \dots\\
\pdimless(\rdimless,\phi) &= - \frac{2}{\hdimless(\REQL)} + \Ca ~\delta p (\rdimless,\phi) + \dots
\end{aligned}
\end{equation}
We wish to determine the leading order corrections to the radius and pressure, i.e.~to find the functions $\delta p (\rdimless,\phi)$ and $f(\phi)$. 

Note that the surfaces are flat in the azimuthal angular direction, $h_\phi=0$, because of the assumption of axisymmetry, so that the gradient of $h$ in the $\xdimless$-direction in the PDE \eqref{eq:ShearPDE} is $\hdimless_{\xdimless}=\hdimless_{\rdimless} \cos\phi$. It is then natural to seek a separable solution of the form $\delta p(\rdimless,\phi) =P(\rdimless)\cos\phi$ with a similar angular dependence for the meniscus position, $f(\phi)=\Rpert\cos\phi$ for some constant $\Rpert$, (i.e.~the meniscus is perturbed a small constant amount, $A \Ca$, in the $x$-direction). From \eqref{eq:ShearPDE}, the equation for the radial dependence of the first-order correction to the pressure is 
\begin{equation} \label{eq:ShearP}
\Ppert'' + \frac{\hdimless+3\rdimless\hdimless'}{\rdimless\hdimless}\Ppert' - \frac{\Ppert}{\rdimless^2} = \frac{6 \hdimless'}{\hdimless^3},
\end{equation}
where we recall that $\hdimless(\rdimless)=1+\rdimless^2/2$. This is subject to two boundary conditions at $\rdimless=\REQL$, where the pressure is set by the curvature through the Young--Laplace equation \eqref{eq:SplittingYL} and there is no fluid flux, i.e.
\begin{align}  
\Ppert(\REQL) &= \frac{2 \hdimless'(\REQL)}{[\hdimless(\REQL)]^2} \Rpert, \label{eq:ShearBVP1} \\
\Ppert'(\REQL) &= \frac{6}{[\hdimless(\REQL)]^2}. \label{eq:ShearBVP2}
\end{align}
The magnitude of the meniscus perturbation $\Rpert$ is unknown, and so an extra boundary condition is required: the pressure must be non-singular at the origin, and considering \eqref{eq:ShearP} at small values of $\rdimless$ gives that
\begin{equation} \label{eq:ShearBVP3}
\Ppert \sim C\rdimless + \frac{3}{8}(2-C)\rdimless^3 \quad \text{as }~ \rdimless\to0
\end{equation} 
for some constant $C$. Note that \eqref{eq:ShearBVP3} is in fact two conditions with an extra unknown $C$: when calculating $\Ppert$, \eqref{eq:ShearBVP3} is imposed a small distance from the origin in both $\Ppert$ and its derivative (so that the unknown $C$ can be determined/cancelled). The problem \eqref{eq:ShearP}--\eqref{eq:ShearBVP3} is therefore fully specified, when given an equilibrium radius $\REQL$.

For a given $\REQL$, the ODE \eqref{eq:ShearP} is solved for $\Ppert(\rdimless)$ subject to the boundary conditions \eqref{eq:ShearBVP2} \& \eqref{eq:ShearBVP3} using MATLAB's in-built BVP solver \texttt{bvp4c}; the relevant radial perturbation $\Rpert$ is then calculated from \eqref{eq:ShearBVP1}. 

To close the problem requires the determination of $\REQL$, which comes indirectly from the dimensionless bridge volume constraint. In practice, it is easier to pick $\REQL$ and find the appropriate dimensionless volume $\Vdimless $ from 
\begin{equation} \label{eq:ShearVol}
\Vdimless = \pi \left( \REQL^2 + \frac{\REQL^4}{4} \right).
\end{equation}

\begin{figure}[tbp]
	\centering
	\includegraphics[width=0.85\linewidth]{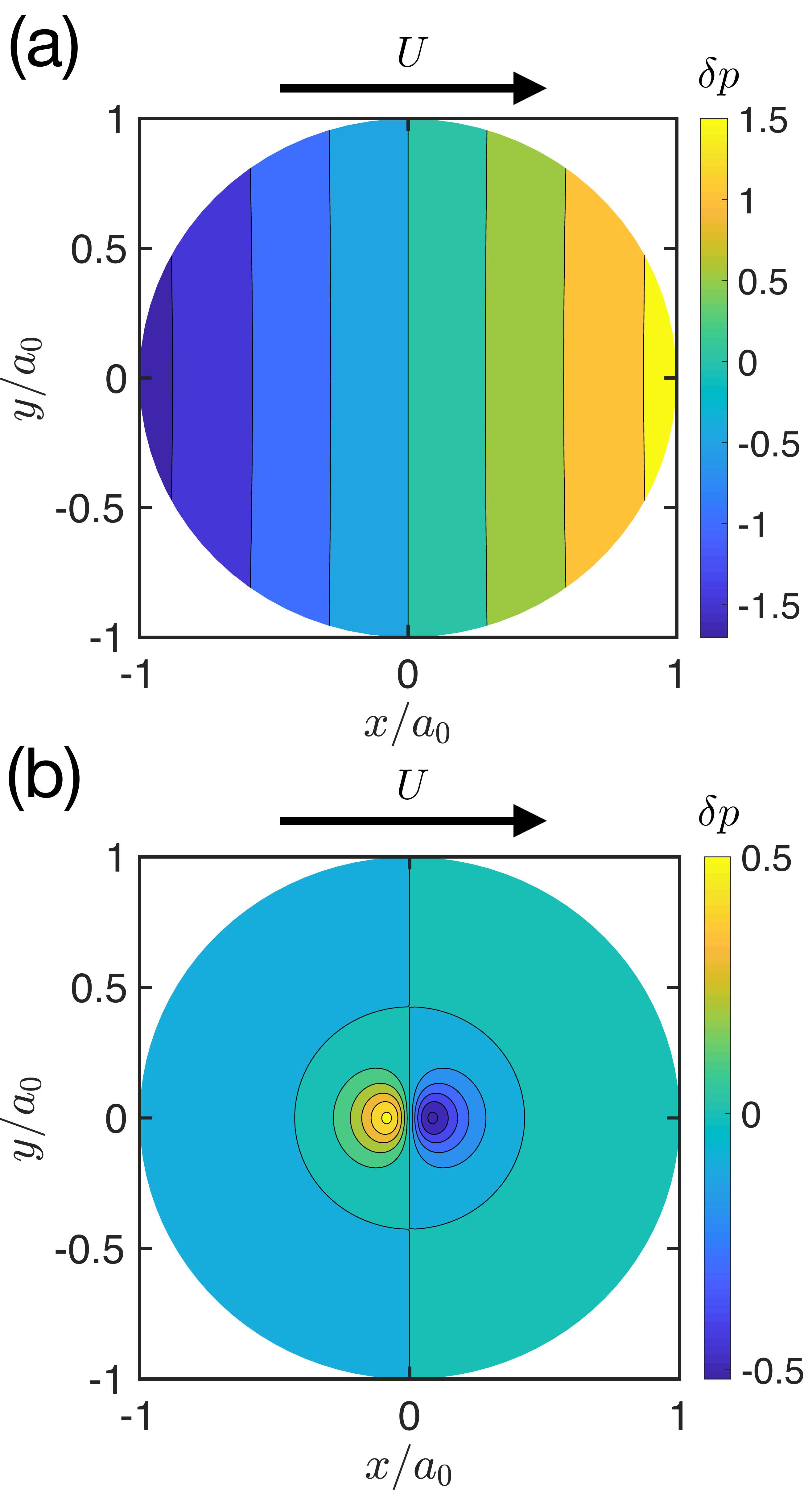}
	\caption[Correction to the liquid pressure during steady shear]{The first order correction to the (dimensionless) pressure field inside the bridge, $\delta p (x,y)=(p-p_0)/(\Ca ~\gamma \cos \theta /\ho)$ as it is sheared in steady state for (a) $V/(R\ho^2)=1$ and (b) $V/(R\ho^2)=100$. The leading order pressure is uniform, $p_0=-2\gbar/h(\Reql)$. Note that the radial lengths have been rescaled so that the equilibrium position has $r=1$ in these images; the corresponding equilibrium radii are $\Reql/\sqrt{R\ho}=0.30$ and $9.5$, respectively. The deviation in meniscus position is not illustrated for simplicity. }
	\label{fig:ShearPressure}
\end{figure}

At leading order in $\Ca$, the pressure is constant and equal to the value in the static case. The change in pressure occurs at $O(\Ca)$; two illustrative examples of the first order correction to the pressure field inside the bridge are shown in fig.~\ref{fig:ShearPressure} for different dimensionless bridge volumes. In fig.~\ref{fig:ShearPressure}a, the dimensionless bridge volume is relatively small and the pressure gradient is approximately uniform and aligned with the direction of the imposed shear. For bridges with larger dimensionless volume, as in fig.~\ref{fig:ShearPressure}b, a recirculating flow is found; this recirculation leads to a much stronger resistive pressure gradient near the origin and a weaker recirculation around it. A natural question to ask is then: how does this affect the resistance to shear? 

\subsection{Resistance to shear during steady motion}

For shearing to be maintained in a steady state, a tangential force must be applied to each solid surface. The dimensional shear force that the liquid applies on the flat plate is $\mathbf{F}_{\text{shear}} = \iint \mu \frac{\partial \mathbf{u}}{\partial z} |_{z=0}~ \mathrm{d}A$, which can be expanded in terms of the perturbation solution and rearranged to give the leading order dimensionless shear force in the $x$-direction
\begin{equation} \label{eq:ShearForce}
\frac{F}{\mu\,U \R} = - 2 \pi \log[\hdimless(\REQL)] - \frac{\pi}{2} \int_0^{\REQL} \rdimless \hdimless \Ppert' + \hdimless \Ppert ~ \mathrm{d}\rdimless.
\end{equation}
The shear force $F$ can be found as a function of the volume $V$, though in practice it is simpler to use the equilibrium bridge radius $\Reql$ as a parameter and compute the appropriate $V$ given by \eqref{eq:ShearVol}, as already discussed. From this shear force, $F[V(\Reql)]$, for a single bridge of volume $V$ the total shear force due to $n$ such bridges can be calculated since $\Ftot (\Vtot)= n F(\Vtot/n)$ with $\Vtot=nV$. 

The contributions to the required horizontal force from both the resulting Couette flow and the counter flow due to the leading order pressure gradient are both linear in the shear velocity $\mathbf{U}=U\mathbf{\hat{x}}$, and so the shear force is linear in $U$. The goal is to find the prefactor for this linear relationship as the bridge volume and geometry are varied and, in particular, to determine the role of the splitting described in \S\ref{sec:RoughSplitting}.

The resistance to shear must be equal and opposite on each of the solids \citep{Batchelor1967}. To corroborate the calculation of the resistance to shear on the planar surface, the force applied on the rough surface has also been calculated in Appendix \ref{app:Shear}, taking careful account of the fact that, in this case, the pressure has a horizontal component when it acts on a curved or sloping surface. The difference in the resistance to shear on each surface is found to be less than 0.1\% over the range presented here, supporting the accuracy of the calculations. Results are therefore presented for the resistance to shearing of the planar surface, but this is equivalent to the shear force experienced by the rough surface. 

\begin{figure}[tbp]
	\centering
	\includegraphics[width=0.9\linewidth]{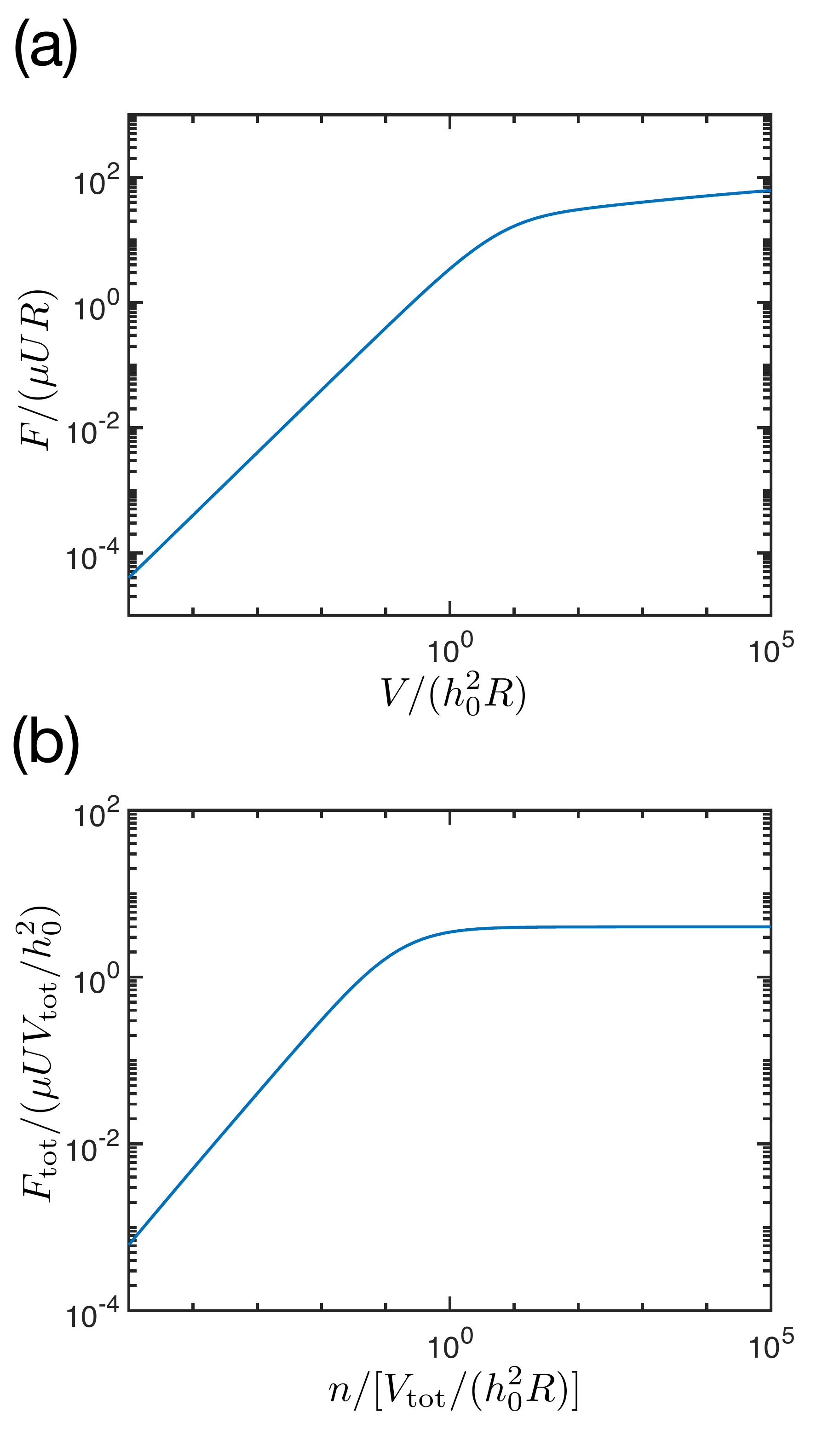}
	\caption[Resistance to shear for a single bridge and multiple bridges]{(a) The magnitude of the dimensionless shear force for a single bridge as a function of the bridge volume. (b) The magnitude of the total shear force when splitting into $n$ bridges. In both cases, results are presented for the shear force applied on the flat plate only; when the results for the shear force on the rough surface are plotted, they are indistinguishable at the scale of the plot.}
	\label{fig:ShearForce}
\end{figure}

The shear force applied on the flat plate by a single bridge depends on which of the two types of pressure fields shown in fig.~\ref{fig:ShearPressure} are observed: these lead to different behaviours of $F(V)$, as can be seen in fig.~\ref{fig:ShearForce}. Figure \ref{fig:ShearForce}a shows the dimensionless applied shear force for a single bridge; in this case, the force increases more rapidly with dimensionless volume when the volumes are small, $V/(\ho^2 \R) < O(10)$, than when they are larger, $V/(\ho^2 \R) \geq O(10)$. For smaller volumes, the shear force increases linearly with volume according to $F/(\mu \, U \R) \approx 4V/(\ho^2R)$ (see Appendix \ref{app:Shear} for a derivation of this relation). For larger bridge volumes, the shear force increases extremely slowly: the dimensionless force $F/(\mu \, U R)$ increases by approximately 10 for each additional factor of 10 increase in the dimensionless volume $V/(\ho^2 R)$ (i.e.~it appears to be approximately logarithmic). The dimensionless shear force in this large bridge volume scenario can be approximated as constant for simplicity, with a value $F/(\mu \, U \R) \approx 100$.

Having considered the force required to shear a plate past a single bridge, it is important to consider the effect of bridge splitting on the total shear force that is generated. For $n$ such bridges with a fixed total liquid volume, $\Vtot$, the total shear force due to this effect is calculated and is plotted in fig.~\ref{fig:ShearForce}b. Again there are two distinct behaviours, transitioning when the number of bridges is in the region of $n \approx 0.1 [\Vtot/(\ho^2 \R)]$. Below this, the total force increases almost linearly with the number of bridges; for a small number of bridges (i.e.~large bridge volumes), the total shear force is approximately independent of the gap width and liquid volume: $\Ftot \approx 100~ \mu \, U \R ~n$.
Above the transition, the dimensionless total force saturates and remains approximately constant as more splitting occurs, giving an upper bound for the shear force that can be achieved by splitting from this mechanism: $\Ftot \approx 4 \mu \, U \Vtot / \ho^2$ (again, see Appendix \ref{app:Shear}). 

Whilst the linearity of the resistance to shear with the sliding velocity found here is similar to that observed in experiments on insects \citep{Federle2004}, the magnitude of the force does not appear to be large enough to explain insects' strong resistance to shear (see Appendix \ref{app:InsectCompare} for a comparison of the calculated shear resistance with measurements on insects). While this hydrodynamic mechanism does not seem to be that behind the resistance to shear seen in these experiments, it is important to emphasize that the experiments on insects do suggest a hydrodynamic mechanism of unknown origin.


\section{Conclusions}


We have considered theoretically the effects of bridge splitting on both normal adhesion and tangential shear forces between rough and flat surfaces. In particular, we have shown that the tendency of capillary bridges to spontaneously migrate to regions in which the gap separation is locally minimal acts to enhance the normal capillary adhesion force, as well as leading to a mechanism through which the shear force is linear in the imposed shearing velocity.

By solving for the shape of capillary bridges numerically, the total adhesive capillary force was calculated as the number of bridges or surface roughness is varied. For perfectly wetting liquids, splitting bridges generally results in an increase in the adhesion force --- an intermediate maximum in the adhesion force is found in this case, corresponding to optimal capillary adhesion away from the very smallest bridges that can exist without rupture. For larger contact angles, the optimal adhesion force is larger and found much closer to rupture, similar to the case of smooth flat surfaces studied previously \citep{DeSouza2008split} (see Appendix \ref{app:Variations}). Fixing the number of capillary bridges, the adhesion is found to increase with the amplitude of the surface roughness, with the maximum capillary adhesion force increasing by orders of magnitude in some cases. 

In the case of a small applied shear, steady solutions to the lubrication equations were found, and the tangential force on the base plate required to maintain a steady state was calculated. This shear force is linear in the imposed velocity and initially increases as the liquid is split into more bridges. This is reminiscent of some experiments on Weaver ants, which found a similar linear behaviour between shear resistance and sliding velocity \citep{Federle2004}, although the mechanism presented in this work does not appear to be large enough to explain the ants' strong resistance to shear (see Appendix \ref{app:InsectCompare}). A potential alternative mechanism that could explain this behaviour is the sliding of a saturated poroelastic medium, such as the recent studies on hydrogels \citep{Ciapa2020,Bonyadi2020,Cuccia2020}; however, a quantification of the magnitude of this effect remains outstanding.

One physical aspect that has been omitted in our models is the possibility of contact line pinning, which could be caused by the presence of small-scale roughness or impurities. Pinning would act to hinder the mechanism for the capillary-induced bridge motion used throughout this work. It is expected that in scenarios where pinning is important, the bridges may not manage to reach the local minima in gap width and so the observed adhesion force would be lower than the values presented here. However, pinning and contact angle hysteresis could contribute additional resistive forces in the steady shear problem (although the combined force from this and our mechanism is unlikely to account for the ant's strong shear resistance).


Our analysis highlights the possible importance of bridge splitting as a means of exploiting surface roughness to enhance capillary adhesion and resistance to shear. In reality, surfaces will not be as idealized as the identical parabolic asperities studied here, but will vary in both wavelength and amplitude. Provided that such a roughness can be characterized well by averages of these properties, our predictions should carry across qualitatively in these more realistic scenarios. However, it is known that the shape of capillary bridges can be affected by the presence of smaller-scale roughness \citep{Wang2009}, and our focus on the behaviour of bridges between one rough and one planar surface may not be appropriate when considering surfaces that mesh well together.
It is hoped that the possible benefits to bridge splitting found here might inspire more detailed experiments with controlled roughness scales that can test these theoretical predictions quantitatively.


\section*{Acknowledgements}

The research leading to these results received funding from EPSRC Grant No. EP/N509711/1 (MB), the Corpus Christi Shand Green-MI Scholarship (MB), the European Research Council under  the European Union's Horizon 2020 Program/ERC Grant 637334 (DV) and a Philip Leverhulme Prize (DV).


\appendix

\section{Splitting perfectly wetting liquid between plates} \label{app:PerfectWetting}


The increase in adhesion force with splitting (and intermediate maxima) seen in fig.~\ref{fig:FlatForce} can largely be explained by a competition between the tension force increasing with $n$ and the decreasing pressure force. However, an increase in adhesion force is also initially seen for perfectly wetting liquid, $\theta=0$, even though in this case there is no tension force, $\FTension=0$. 

Heuristically, this is because of changes in shape of the bridge as the liquid is divided more: when splitting, liquid is initially lost preferentially from the centre, or `neck', of the bridge. This means that the wetted area does not decrease in magnitude as much as might have been expected when splitting perfectly cylindrical bridges (in which the wetted area of a single bridge $A_{\text{wet}} \propto \V \propto 1/n$). Despite this `necking' also causing the curvature to become more positive (i.e.~a more repulsive pressure), the pressure force-per-bridge (the product of curvature and wetted area) decreases slowly enough with number of bridges that having more bridges still results in a larger total adhesion force. 

As the bridges are split further, the wetted area begins to decrease (and curvature to increase) more rapidly with $n$, causing an eventual decrease in the pressure force, and hence total force. 
This increase in the pressure force is only seen when the liquid is very wetting; for contact angles as small as $\theta=30^\circ$ the pressure force decreases with increasing number of bridges and the increase in adhesion with splitting is simply due to the tension force only.

\section{Rough capillary adhesion with other wettabilities and roughness geometries} \label{app:Variations}

In this appendix, we consider modifications to the results presented in the main text when taking account of different liquid wettabilities and roughness geometries. 

\subsection{Variation with contact angle}

\begin{figure}[tbp]
	\centering
	\includegraphics[width=0.85\linewidth]{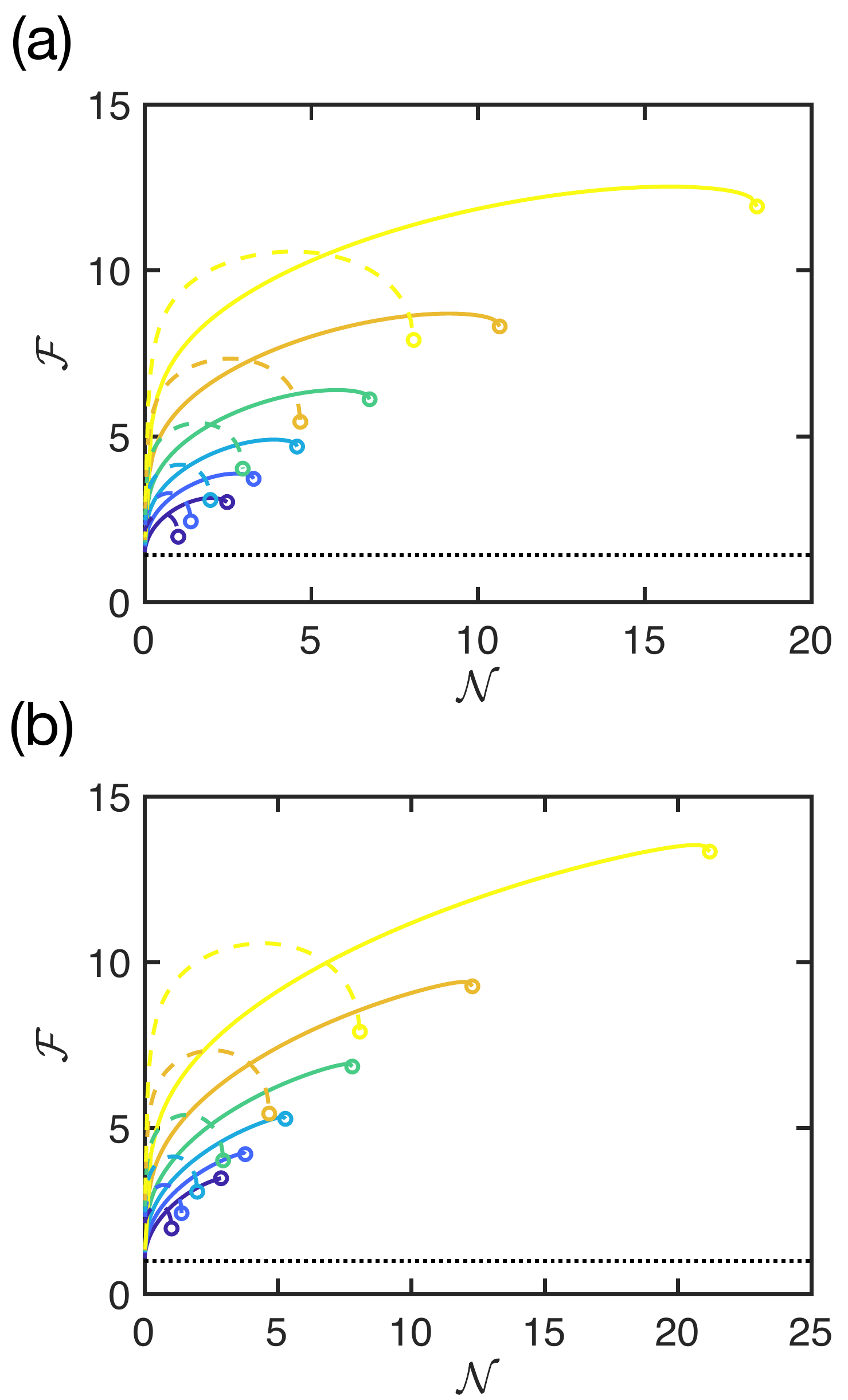}
	\caption[Total adhesion force for different contact angles]{Effect of varying contact angle $\theta$: the dimensionless total adhesion force $\Fbridge$ as a function of the scaled number of bridges $\nbridge$ for the roughness amplitudes $\Dh/\h=\{0,0.1,0.2,0.3,0.4,0.5\}$ when (a) $\theta=45^\circ$, and (b) $\theta = 60^\circ$. In both cases, the dashed curves indicate the results for a perfectly wetting bridge, $\theta=0$, for comparison with the results in the main text while the dotted lines show the simple parallel plates model, $\Fbridge=2\cos\theta$. In both (a) and (b), the direction of increasing $\Dh/\h$ and colour code are the same as in fig.~\ref{fig:RoughForce}a. }
	\label{fig:VaryCA}
\end{figure}

Results in the main text were presented for perfectly wetting liquid bridges with contact angle $\theta=0$. If the contact angle, $\theta$, is varied, then the results remain qualitatively similar (see fig.~\ref{fig:VaryCA}). Similarly to the perfectly wetting bridges, rupture is found to occur later and the adhesion force is seen to increase as the roughness amplitude is increased. In fact, splitting is even more beneficial at these contact angles than for perfect wetting
: a similar magnitude (or larger) force can be obtained by splitting, but this can be many times the force found in the large single bridge case of eqn.~\eqref{eq:FlatCase}, due to the factor $\cos \theta$ there. Additionally, it is possible to get many more bridges before rupture occurs. However, it must be noted that as the contact angle increases, the maximum moves closer to rupture and is not as broad --- a more accurate (discrete) choice of $n$ may be required to maximize the force whilst avoiding rupture. 


\subsection{Variation with aspect ratio}

\begin{figure}[tbp]
	\centering
	\includegraphics[width=0.85\linewidth]{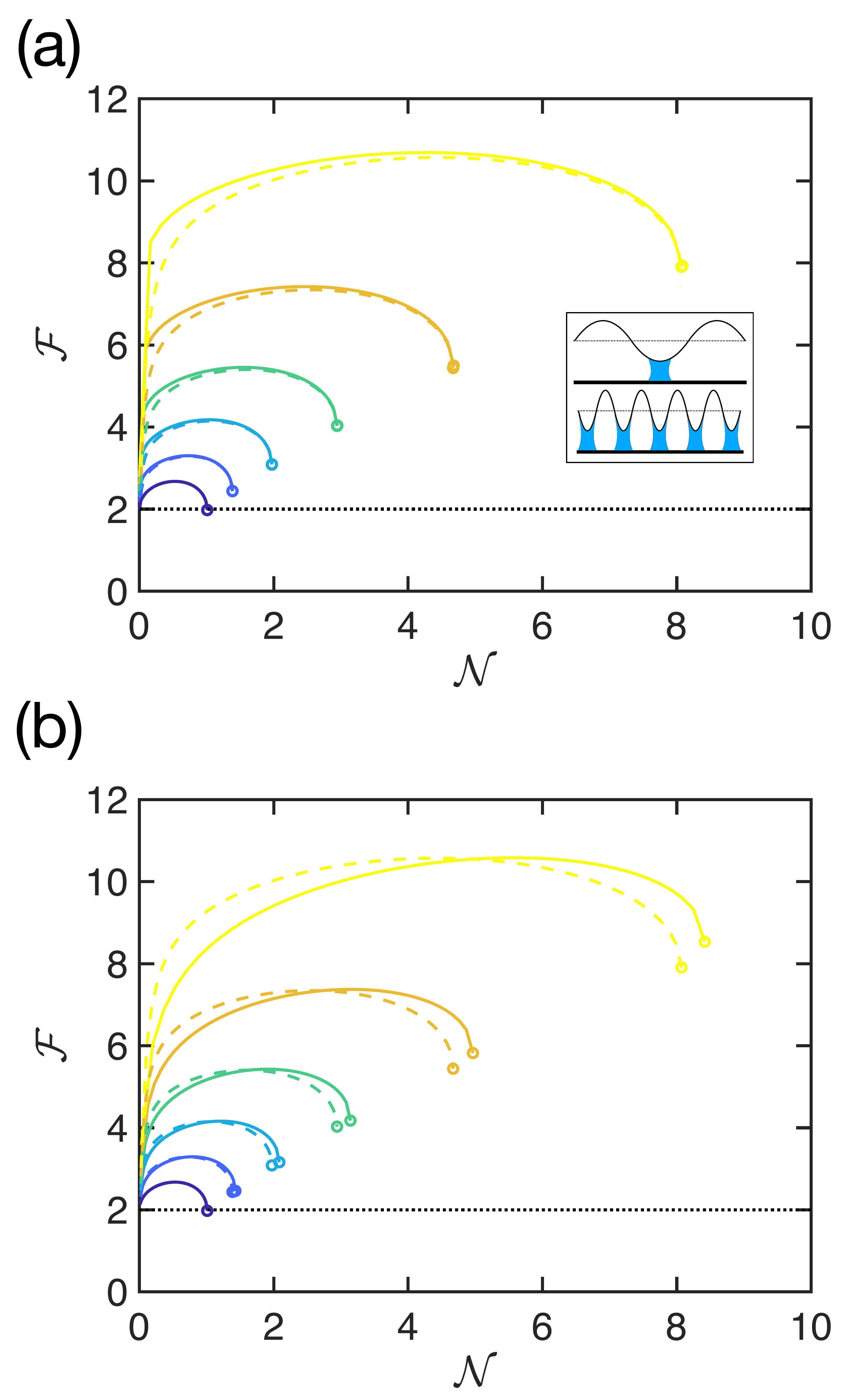}
	\caption[Total adhesion force for different aspect ratios and a conical roughness]{ Effect of varying the aspect ratio and roughness type: the dimensionless total adhesion force, $\Fbridge$, as the scaled number of bridges $\nbridge$ is varied for various fixed roughness amplitudes $\Dh/\h=\{0,0.1,0.2,0.3,0.4,0.5\}$ when the liquid is perfectly wetting, $\theta=0$. 
	(a) Varying $\asp$: solid curves show numerical results with $\asp=100$, while dashed curves reproduce the results of  fig.~\ref{fig:RoughForce}a (for $\asp=10$). Inset: representative profiles showing rough surfaces with aspect ratios of $\asp=10$ (lower) and $\asp=100$ (upper). 
	(b) The conical asperity case: solid curves show results for the conical aspect ratio $\beta=\sqrt{20}$ are compared to the parabolic case with $\asp=10$ (dashed).  
	In both (a) and (b) the direction of increasing $\Dh/\h$ and colour code are the same as in fig.~\ref{fig:RoughForce}a, and the dotted lines again show the simple parallel plates model, $\Fbridge=2$, from eqn.~\eqref{eq:FlatCase}.} 
	\label{fig:Asp_Sawtooth}
\end{figure}

To investigate how the results change as the roughness aspect ratio is altered, $\asp$ (given by eqn.~\eqref{eq:AspectRatio}), the force at fixed roughness amplitude is plotted when $\asp=100$ and compared to the values obtained at $\asp=10$ (the value used for results in the main text). This is shown in fig.~\ref{fig:Asp_Sawtooth}a. If the average height, $\h$, is held constant and the same roughness amplitudes, $\Dh$, are compared then increasing $\asp$ corresponds to increasing the radius of curvature of the asperity, $\R$, i.e.~making the roughness flatter. At the larger aspect ratio, the force initially increases more rapidly with $n$, but otherwise the behaviour is qualitatively the same: there is an intermediate maximum in the force when varying $n$ and the force increases with the roughness amplitude. There is minimal change in both the force and the value of $n$ at which the bridge ruptures, varying by only a few percent at most.

From this comparison, it seems that changing the value of $\asp$ does not change the results a great deal. However, it should be noted that larger $\asp$ may mean a sparser packing of asperities (since the radius of curvature $\R$ is larger, see inset to fig.~\ref{fig:Asp_Sawtooth}a), and so a larger adhesive area may be needed to generate the same absolute adhesion force. 

\subsection{Geometric variation: the conical case}

Here, we investigate the effect of changing the local geometry of the roughness. If the roughness is not sufficiently smooth then locally it may look more conical in nature. 

Consider a roughness that is locally conical, i.e.~$h(r) = \ho + B r$, for some slope $B$. The aspect ratio $\beta = \Dh/B \h$ is fixed in the same manner that in the parabolic case $\alpha$ was fixed. The eqns.~\eqref{eq:BVP} are the same as for the parabolic case, but the boundary conditions \eqref{eq:BCs} are altered appropriately.

An example of the total adhesion force is given in fig.~\ref{fig:Asp_Sawtooth}b for the case $\beta=\sqrt{20}$. This value of $\beta$ is used to compare to the $\asp=10$ case from the main text, because when $\beta=\sqrt{2\asp}$ then the radius at which $\xi(1)=1$ (i.e. the $r$ at which $h(r)=\h$) is the same in both cases, and so their aspect ratios can be considered equivalent.

Again, the results are qualitatively similar to those presented in the main text: there is an intermediate maximum and the force increases with roughness amplitude. It can be concluded that, regardless of whether the roughness looks locally parabolic or conical, splitting can give a significant increase in adhesion force. 

\section{Calculation of the resistance to shear} \label{app:Shear}

\subsection{Shear force on the rough surface}

In \S\ref{sec:ShearSplitting} results were presented for the shear force applied by the liquid on the flat surface. Here, the shear force on the rough surface is calculated. In this case, it is important to account for the fact that (in addition to the usual viscous shear $\mu \, \partial \mathbf{u}/\partial z$) the pressure can have a horizontal component that modifies the resistance to shear: the rough surface is sloped and so its normal action leads to a small horizontal component. 
In terms of the small $\Ca$ expansion, this pressure contribution to the shear comes in two main parts, both of which give a leading order correction to the shear that is not seen in the flat case: 
(i) the uniform leading order pressure $-2/\hdimless(\REQL)$ acting over its perturbed radius, and
(ii) the asymmetric pressure correction $\delta p$ acting over the equilibrium footprint $\REQL$. 

The leading order dimensionless shear force that the liquid applies on the rough surface in the $x$-direction is found to be 
\begin{multline} \label{eq:RoughShear}
\frac{F}{\mu \, U \R} = 2 \pi \log[\hdimless(\REQL)] + \frac{2 \pi \REQL^2}{\hdimless(\REQL)} \Rpert \\
- \frac{\pi}{2} \int_0^{\REQL} \rdimless \hdimless \Ppert' + \hdimless \Ppert + 2 \rdimless^2 \Ppert~ \mathrm{d}\rdimless
\end{multline}
which can be compared to the shear force on the plane \eqref{eq:ShearForce}.
The second term here is the contribution caused by the moving edge of the liquid bridge; the final term in the integral is the part from the asymmetric pressure correction.

We find that the calculated shear force on the rough and planar surfaces differ by less than $0.1\%$ over the range considered --- this difference is indistinguishable at the scale of fig.~\ref{fig:ShearForce}; default tolerances were used when implementing \texttt{bvp4c} (relative and absolute tolerances were $10^{-3}$ and $10^{-6}$, respectively). The difference between shear forces observed here may therefore be attributed to numerical error since the maximum error is within the requested error tolerance. It is therefore suggested that the shear force on each surface is the same, and focus in the main text only on the shear force on the planar surface for simplicity.


\subsection{Behaviour for small bridges}

If the bridges are small (but still remain in the lubrication limit), $\REQL \ll 1$, then the rough surface will look flat on the scale of the bridge with $\hdimless \approx 1$ (and also $\log [\hdimless(\REQL)] \approx \REQL^2/2$). The solution of \eqref{eq:ShearP} is then
\begin{equation}
\Ppert(\rdimless) \approx 6 \rdimless, 
\end{equation}
because the condition \eqref{eq:ShearBVP3} enforces that the pressure behaves linearly at small radii, and the no flux boundary condition at the edge, \eqref{eq:ShearBVP2}, gives a pressure gradient of 6. This linearity in $\Ppert$ can be seen in fig.~\ref{fig:ShearPressure}a. The shear force on the flat surface can then be approximated by
\begin{equation}
\frac{F}{\mu \, U \R} \approx - \pi \REQL^2 - \frac{\pi}{2} \int_0^{\REQL} 12 \rdimless ~ \mathrm{d}\rdimless \approx - 4 \pi \REQL^2.
\end{equation} 
Furthermore, in this small bridge case, the dimensionless bridge volume is $\Vdimless \approx \pi \REQL^2$ so that the predicted shear force behaviour is
\begin{equation}
\frac{F}{\mu \, U \R} \approx -4 \frac{V}{\ho^2 \R}, \qquad \text{ or } \qquad \frac{\Ftot}{\mu \, U \Vtot / \ho^2} \approx -4
\end{equation}
which is observed in the results of fig.~\ref{fig:ShearForce} for small $V$ or large $n$.

\section{Resistance to shear in insects} \label{app:InsectCompare}


\begin{figure}[tbp]
	\begin{center}
		\includegraphics[width=\linewidth]{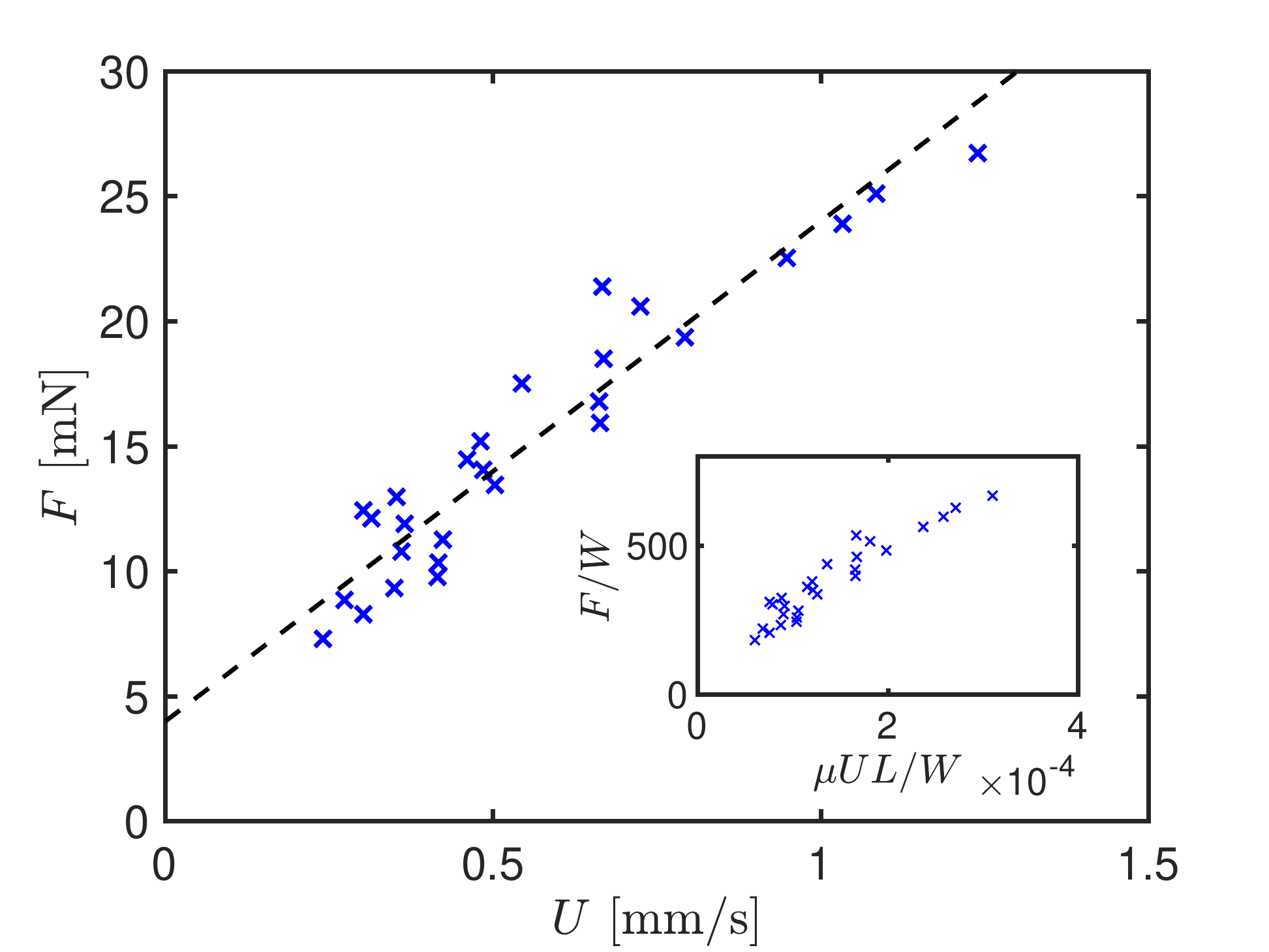} 
		\caption[Data from experiment of an ant on a turntable]{Resistance to shear provided by an Asian Weaver ant. Experimental data from \citet{Federle2004} are digitally captured and plotted in the main figure. The data appear to be well-approximated by a linear relationship $F=20U+4$ (black dashed line) with $F$ measured in mN and $U$ in mm/s. Inset: An approximation for the data as would be seen on a Stribeck curve. For this inset, the load was estimated as $W=40\mathrm{~\mu N}$, the length of contact $L=0.1\mathrm{~mm}$ (both from data in \citep{Federle2004}) and the dynamic viscosity $\mu=0.1\mathrm{~Pa\,s}$  \citep{Federle2002}.}
			\label{fig:TurntableExpt} 
	\end{center}
\end{figure}

When insects adhere to a surface they are able to withstand large normal loads relative to their body weight, but also comparatively large shear forces \citep{Dirks2014,Dirks2011review}; a long-standing problem in insect adhesion is to understand the origins of this resistance to shear. Experiments on Asian Weaver ants suggest that the shear force increases linearly with the velocity of shearing \citep{Federle2004}, with only a relatively small threshold at $U=0$ (see fig.~\ref{fig:TurntableExpt}). This linear relationship between force and velocity is evocative of the relationship between rate of shearing and shear stress acting on two parallel plates separated by a Newtonian liquid \citep{Batchelor1967,Worster2009}. 

A common means of plotting such data is via a Stribeck curve \citep{Hamrock2004} showing friction coefficient as a function of dimensionless speed --- this clearly distinguishes between boundary lubrication (that is independent of speed) and hydrodynamic lubrication (that increases linearly with speed). 
The inset of fig.~\ref{fig:TurntableExpt} shows an estimated Stribeck curve (based on an assumed load $W=40\mathrm{~\mu N}$, contact length $L=0.1\mathrm{~mm}$ and dynamic viscosity $\mu=0.1\mathrm{~Pa\,s}$), where friction coefficients are plotted as a function of the Hersey number ($\mu \,UL/W$, which is a dimensionless lubrication parameter but for our purposes can be thought of as a dimensionless speed). Again, this shows an approximately linear trend, suggesting the origin of this shear force is hydrodynamic.
However, a simple model of a Newtonian liquid bridge between plates does not appear to provide a large enough resistance to shear to explain the observed forces \citep{Federle2002}; further, a steady state cannot exist in the case of a finite liquid bridge, which would be smeared over the plate by the shearing.

One possible clue as to insects' strong resistance to shear is the presence of emulsion droplets in their oily secretion. As well as the potential benefit of enhancing capillary adhesion, it has also been suggested that these emulsion droplets could endow the liquid with non-Newtonian properties, such as a yield stress, thereby also allowing for resistance to shear \citep{Dirks2010,Drechsler2006}. Another possibility is that contact angle hysteresis and contact line pinning may oppose shearing as the pull of the contact line contributes a horizontal component \citep{Dirks2014}. However, both of these mechanisms predominantly generate a static resistance to shear and cannot explain the simple linear relationship between the resistance to shear and the imposed shear velocity that is observed experimentally (fig.~\ref{fig:TurntableExpt}). 

Our theory for shear resistance due to steady capillary migration on rough surfaces (presented in \S\ref{sec:ShearSplitting}) does show a linear dependence between the shear force and velocity, just as observed by \citet{Federle2004} for Asian Weaver ants on a rotating turntable. Their data suggest a linear relationship with a gradient close to ${c=F/U \approx 20 \mathrm{~N\,s/m}}$, as shown in fig.~\ref{fig:TurntableExpt}. The question is then to explain the order of magnitude of the friction constant $c$ --- can it be explained as being the result of the splitting of capillary bridges on a rough surface?

\subsection{Comparison of shear resistance}


To make a quantitative comparison between the predictions of our model and the experimental observations of \citet{Federle2004}, we take published values of parameters for Asian Weaver ants from \citet{Dirks2011secretion,Federle2004}, \citet{Federle2002} and \citet{Dirks2010}. Since the mechanism for generating a shear resistance discussed here would work equally well for a single curved object (i.e.~the whole foot pad, or a single asperity,) as many smaller ones, it is not immediately clear whether the relevant scale here is the whole oil secretion (volume $V\approx 10^3 \mathrm{~\mu m}^3$ \citep{Dirks2011secretion} with radius of curvature $\R\approx100\mathrm{~\mu m}$ of the whole footpad \citep{Federle2004}) or the emulsion droplets (smaller volume and a radius of curvature from local asperities). In either case, the experimental capillary number is small, $\Ca \lesssim 0.1$ 
(based on a shearing rate $U \approx 1 \mathrm{~mm/s}$, viscosity $\mu \approx 0.1\mathrm{~Pa\,s}$ \citep{Federle2002}, and assuming a similar surface tension to other insect species $\gamma \approx 20$--$30 \mathrm{~mN/m}$ \citep{Dirks2014,Gernay2016}), provided that the separation $\ho > 100~\mathrm{nm}$. 
Hence the small capillary number theory developed here is expected to be valid for these experiments. 

At the scale of the whole foot, a single bridge consisting of the entire oily secretion may be considered, and results applied from fig.~\ref{fig:ShearForce}a. If the ant's foot is reasonably far from the surface, $\ho > \sqrt{V/10\R} \approx 1 \mathrm{\mu m}$, then the force would be expected to be in the linear regime with friction constant $c=F/U \approx 4 \mu V/\ho^2$. Using the previously mentioned values, the friction coefficient is found to be $c <10^{-4} \mathrm{~N\,s/m}$. This is around five orders of magnitude smaller than the observed value. If the separation were much smaller than this, $\ho<1\mathrm{~\mu m}$, then instead the force-velocity relationship would be expected to saturate, so that $c\approx 100 \mu R \approx 10^{-3} \mathrm{~N\,s/m}$ --- again, much smaller than observed experimentally. 

Alternatively, it is possible to estimate the shear force from bridges on the scale of the watery droplets within the emulsion, using results from fig.~\ref{fig:ShearForce}b. Suppose that the secretion consists of 10\% watery droplets by volume, of typical extent $\Reql\approx1\mathrm{~\mu m}$ (estimated from images of the insect secretion \citep{Dirks2010}), and that these all form bridges on a substrate roughness of a similar scale, $R \approx 1 \mathrm{\mu m}$. Then, since the radial lengthscale is $1/100$th of the whole foot and the volume is $1/10$th of the whole secretion, approximately $n \approx 1000$ bridges would be expected for the same separation $\ho$. At the scale of the individual bridges, the insect foot will look flat compared to the substrate roughness. It is then reasonable to consider a shear being applied on the flat surface, whilst measuring the resulting resistance to shear. If the surface separation is large enough $\ho > \sqrt{0.1 \Vtot/ n \R} \approx 0.1 \mathrm{~\mu m}$, then the shear friction is bounded above by $c \approx 4 \mu \Vtot / \ho^2 < 10^{-3} \mathrm{~N\,s/m}$; if $\ho$ is smaller than this then the friction will be $c \approx 100 \mu \R n \approx 10^{-2} \mathrm{~N\,s/m}$. Again, this remains significantly smaller than the value measured experimentally. 

It can therefore be concluded that the shear resistance provided by the migration of capillary bridges is not sufficient to generate the  shear forces observed in insects (even after accounting for the fact that the insect has 6 points of contact). Nevertheless, this does give an example of a mechanism that would yield a linear relationship between shear force and velocity. Additionally, this bridge migration may help to retain liquid during shear, when in other scenarios the bridges may be sheared away or end up coating the surfaces. 



\bibliographystyle{unsrtnat}
\bibliography{RoughAdhesion}

\end{document}